\documentclass[twocolumn,aps,prd,showpacs,showkeys,nofootinbib]{revtex4}
\usepackage{amsmath}
\usepackage{amssymb}
\usepackage{graphicx}
\usepackage{hyperref}
\hypersetup{colorlinks=true,citecolor=blue, linkcolor=red, urlcolor=magenta, hidelinks}

\begin{document}

	\title{Non-parametric reconstruction of interaction in the cosmic dark sector}
	
	\author{Purba Mukherjee$^1$} 
	\email{pm14ip011@iiserkol.ac.in} 
	
	\author{Narayan Banerjee$^2$}
	\email{narayan@iiserkol.ac.in}

	\affiliation{\\~\\$^1$$^{,2}$Department of Physical Sciences,~~\\Indian Institute of Science Education and Research Kolkata,\\ Mohanpur, West Bengal - 741246, India.}

\begin{abstract}
The possibility of a non-gravitational interaction between the dark matter and the dark energy has been reconstructed using some recent datasets. The crucial aspect 
is that the interaction is not parametrized at the outset, but rather reconstructed directly from the data in a non-parametric way. The Cosmic Chronometer Hubble data, 
the Pantheon Supernova compilation of CANDELS and CLASH Multy-Cycle Treasury programs obtained by the HST, and the Baryon Acoustic Oscillation Hubble data have been considered
in this work. The widely accepted Gaussian Process is used for the reconstruction. The results clearly indicate that a no interaction scenario is quite a possibility. Also, 
the interaction, if any, is not really significant at the present epoch. The direction of the flow of energy is 
clearly from the dark energy to the dark matter which is consistent with the thermodynamic requirement.
\end{abstract}

\vskip 1.0cm



\maketitle

\section{Introduction}

That the universe is expanding in an accelerated fashion is now believed to be a certainty. Albeit the indication started pouring in more than twenty years 
ago\cite{reiss, perl}, there is yet to be any unequivocally accepted theoretical framework that can settle this puzzle. Surely the good old cosmological constant 
$\Lambda$, the ``weight of the vacuum'' as described by Padmanabhan, can resolve the issue\cite{padma}, but it has its own share of problems, such as the 
enormous mismatch between the theoretically predicted value and observationally required one. This discrepancy is quite elaborately discussed in the reviews 
\cite{padma} and \cite{peebles}. Various inconsistencies with $\Lambda$ were in fact known even before its requirement as dark energy, the driver of 
the acceleration of the universe. An account of this can be found in the famous work of Weinberg\cite{wein}. \\

As the easiest choice of $\Lambda$ as the ``dark energy'' runs into trouble, many other alternative models were suggested starting from the introduction of exotic fields
of various forms in the energy distribution in the realm of general relativity (GR), to different kinds of modification of GR as the theory of gravity. For a recent review,
we refer to the work of Brax\cite{brax}. \\

In various dark energy models, the normal practice is to assume that the exotic dark energy and the familiar cold dark matter evolve independent of each 
other. For early reviews, we refer to \cite{padma}, \cite{sami} and \cite{varun}. However, the so-called coincidence problem\cite{zlatev}, which poses the question 
why the dark energy and the dark matter densities are of the same order of magnitude at the present epoch, inspired a search for any non-gravitational interaction 
between them. In fact a possible interaction between the vacuum energy and the pressureless matter was suggested long back, for example, we refer to the 
investigations by Henriksen\cite{henrik}, and Olson and Jordan\cite{olson}. \\

In the context of the present acceleration of the universe, the idea is that the dark matter and the dark energy may not evolve independently, there is rather a transfer of 
energy between them. As a result, they do 
not satisfy individual conservation equations and only the total energy is conserved via the equation 
$\nabla_\mu (T_m^{\mu \nu} + T_D^{\mu \nu} ) = 0$, where $T_m^{\mu \nu}$, $T_D^{\mu \nu}$ denote the energy-momentum tensors for the dark matter and the dark energy 
respectively. A lot of investigation in this direction has already been carried out, we cite \cite{farrar, caiwang, amendola, guo, he, caldera, cai_su, honorez, besseda, yang14} 
to name a few. A function $Q$ is introduced 
such that, $\nabla_\mu T_m^{\mu \nu} = -\nabla_\mu T_D^{\mu \nu} \propto Q$. The function $Q$ clearly determines the rate of transfer of energy from one sector to the other. As 
the origin and nature of the non-gravitational interaction is not known, $Q$ is phenomenologically chosen in various forms. Usually it is assumed to be proportional to the Hubble 
parameter $H$ and one of the densities that of the dark matter or the dark energy. More general forms, where a linear combination of both the densities or even their derivatives 
are involved can also be found in the literature. For various choices and their consequences in the context of observations, we refer 
to the references \cite{r31,r32,r33,r34,holo_ankan,r35,r36,ankan_pan,holo_purba,r37,r38,r39,r40,r41,r42,r45,r47,r48,r49,r50,r52}. It was shown that the 
interacting scenario can also potentially take care of the issues connected to the local value of the Hubble 
parameter\cite{r40,r50,r53}. For an exhaustive review, the work of Wang, Abdalla, Atrio-Barandela and Pav\'{o}n \cite{Q_review} is quite useful.\\

A reconstruction of the interaction from observations normally depends on the parametric form of $Q$, and the parameters are estimated from  
datasets. Another approach for reconstruction is the non-parametric one where no particular functional form of $Q$ is assumed. Rather, one makes an attempt to ascertain the 
quantity directly from data. So this is clearly more unbiased. The non-parametric approach finds a lot of application in cosmology. It all started with the reconstruction of 
physical quantities like the equation of state parameter of the dark energy, the scalar potential when a quintessence scalar field is used as the dark energy 
etc.\cite{sahl1, sahl2, holsclaw1, holsclaw2, holsclaw3, critt, sanjay, zzhang}. Recently the non-parametric approach finds relevance in the reconstruction of kinematical 
quantities like the deceleration parameter $q$\cite{bilicki,lin,xia,carlos,nunes,jesus_nonpara,arjona,purbaq} and the jerk parameter $j$\cite{purba}. \\

Given that there is no \textit{a priori} reason to rule out an interaction in the dark sector on one hand, and the lack of any compulsive theoretical model for that on the other, 
certainly a reconstruction of the interaction $Q$, in an unbiased way without assuming any functional form of $Q$, deserves a lot more attention than that is available 
in the literature. Cai and Su\cite{cai_su} investigated the possible interaction, in a way independent of any specific form, by dividing the whole range of redshift into a few 
bins and setting the interaction term a constant in each redshift bin. The result indicated that there could be an oscillatory interaction. Wang \textit{et al.}\cite{wang_pca} 
adopted a non-parametric Bayesian approach, and indicated that an interacting vacuum is not preferred over the standard $\Lambda$CDM. \\

Yang, Guo and Cai \cite{yangguocai} presented a non-parametric reconstruction of the interaction between dark energy and dark matter directly from SNIa Union 2.1 data using 
the Gaussian Process. They found that unless the equation of state (EoS) parameter $w$ for the dark energy deviates significantly from $-1$, the interaction is not evident. If 
$w$ is widely different from $-1$, the interaction cannot be ruled out at a $95\%$ confidence level. A recent analysis, also another non-parametric 
reconstruction using 
the Gaussian Process, indicates a very interesting possibility that a gravity wave signal might carry signatures of the interaction in the dark sector encoded in the wave signal. 
This work by Cai, Tamanini and Yang \cite{Qlisa} is based on LISA space-based interferometer. It was shown that a 10 year survey can unveil the interaction in a wide redshift 
domain between $1<z<10$. \\

The motivation of this work is to undertake a non-parametric reconstruction of the interaction term $Q$ as a function of the redshift $z$ using recent cosmological datasets. 
The Cosmic Chronometer Hubble data, the Pantheon Supernova compilation of CANDELS and CLASH Multy-Cycle Treasury programs obtained by the HST, and the Baryon Acoustic 
Oscillation Hubble data are utilized for the purpose. The method adopted is the widely used Gaussian Process. We consider three cases of the dark energy equation of state (EoS), 
the decaying vacuum energy $\Lambda$ with $w = -1$, the $w$CDM model and the Chevallier-Polarski-Linder (CPL) parametrization of dark energy. Here $w$ is the EoS parameter. 
For different combinations of datasets and for different choices of dark energy, the most important common feature found is that ``no interaction'' is almost always included 
in $2\sigma$ and definitely in $3\sigma$. 
Also, at the present epoch interaction is hardly indicated,  
if no interaction is beyond $2\sigma$, that would have been in the past, close to or beyond $z\sim 0.5$. \\

This manuscript is organized as follows. In section II, we describe the model with an interaction in the dark sector. Section III deals with the results of the actual 
reconstruction following a brief description of the Gaussian Process, a summary of the datasets utilized and the methodology.  Fitting functions 
for the reconstructed interaction are shown in Section IV, for different combinations of datasets. In section V, a discussion on the evolution of the density parameters 
is included. In section VI, attempts have been made for testing the reconstructed model against the laws of thermodynamics. In the last section, we discuss the 
results obtained and make some concluding remarks.

\section{The Model}

The infinitesimal distance element in a spatially flat, homogeneous and isotropic universe is given by the Friedmann-Lema\^{i}tre-Robertson-Walker (FLRW) metric
\begin{equation}
ds^2 = -c^2 d t^2 + a^2(t) \left(d r^2 + r^2 d\theta^2 + r^2 \sin^2\theta d\phi^2 \right),
\end{equation} where $a(t)$ is the scale factor. \\

The Hubble parameter, defined as the expansion rate of the universe, is given by
\begin{equation}
H = \frac{\dot{a}}{a},
\end{equation}
where a `dot' signifies derivative $w.r.t.$ the cosmic time $t$. This expression can again be written as a function of the redshift $z$, given by the relation 
$1+z = \frac{a_0}{a}$ where any subscript $0$ indicates the present value of the quantity considered and $a_0$ is taken to be unity. We define the dimensionless Hubble parameter as, 

\begin{equation}
E(z) = \frac{H(z)}{H_0}.
\end{equation} 

In a flat universe with an interaction between dark energy and dark matter, the Einstein equations describing the evolution of our universe given by,
\begin{eqnarray}  
H^2 &=& \frac{8 \pi G}{3}(\rho_m + \rho_D) \label{friedmann},\\
\dot{H} + H^2 &=& -\frac{8 \pi G}{6}(\rho_m + \rho_D + 3 p_D ),
\end{eqnarray} 
where $\rho_m$ denotes the energy density of dark matter, $\rho_D$ the energy density of dark energy component and we consider $8 \pi G = 1$; $p_D$ signifies 
the pressure component from the dark energy sector, and $p_m = 0$ for pressureless dust. The energy conservation equation is given by the contracted Bianchi 
identity,
\begin{equation}
\dot{\rho} + 3H (1+w_{\mbox{\tiny eff}})\rho  = 0,
\end{equation} 
where the total energy density is given by, $\rho = (\rho_m + \rho_D)$ and $w_{\mbox{\tiny eff}}$ is the effective equation of state defined as 
\begin{equation}
w_{\mbox{\tiny eff}} = \frac{p}{\rho} = \frac{p_D}{\rho_m + \rho_D}.
\end{equation} 
However, this conservation equation can be separated into two parts,
\begin{eqnarray} 
\dot{\rho_m}+ 3H\rho_m &=& -Q ,\\
\dot{\rho_D} + 3H (1+w)\rho_D &=& Q,
\end{eqnarray} 
where, $w = \frac{p_D}{\rho_D}$ is the equation of state of DE and $Q$ describes the rate of transfer of energy between dark matter and dark energy. When 
$Q = 0$ and $w = -1$, one recovers the standard $\Lambda$CDM model. Unlike the usual practice of parametrizing the interaction term $Q$ using any parametric form 
proportional to $H\rho$, here we want to reconstruct it directly from observational data using a non-parametric method.\\

By using $\frac{d}{dt} \equiv - H(1+z) \frac{d}{dz}$, one can rewrite the conservation equations, with the redshift $z$ as the argument, in the form
\begin{eqnarray} \label{conservation}
-H (1+z) \rho_m^\prime + 3 H \rho_m &=& -Q (z), \\
-H (1+z) \rho_D^\prime + 3 H (1+w) \rho_D &=& Q (z).
\end{eqnarray}

Equation \eqref{friedmann} can be written in terms of the dimensionless Hubble parameter $E = \frac{H}{H_0}$ as, 
\begin{equation} \label{fried_reduced}
E^2(z) = \tilde{\rho}_m + \tilde{\rho}_D,
\end{equation} 
where,  $\tilde{\rho}_m$ and $ \tilde{\rho}_D$ are $\rho_m$ and $\rho_D$ scaled by a factor of $\frac{1}{3 H_0^2}$ respectively. On differentiating 
Eq. \eqref{fried_reduced} w.r.t. $z$, we get
\begin{equation} \label{fried_diff}
2 E E'(z) = \tilde{\rho}_m^\prime + \tilde{\rho}_D^\prime .
\end{equation} 

Both the conservation equations \eqref{conservation} can be reduced to the following forms,
\begin{eqnarray} 
-E (1+z) \tilde{\rho}_m^\prime + 3 E \tilde{\rho}_m &=& -\tilde{Q} (z), \label{cons_reduced_m}\\
-E (1+z) \tilde{\rho}_D^\prime + 3 E (1+w) \tilde{\rho}_D &=& \tilde{Q} (z). \label{cons_reduced_d}
\end{eqnarray} 
The dimensionless $\tilde{Q}$ characterizes the interaction, where $\tilde{Q} = \frac{1}{ 3 H_0^3} Q$.\\

By combining equations \eqref{fried_diff} with \eqref{cons_reduced_m}-\eqref{cons_reduced_d}, one can obtain,
\begin{widetext}
	\begin{equation} \label{Q_h}
	\tilde{Q} = \left( \frac{E^2(1+w)}{w} + \frac{(1+z)E^2 w'}{3 w^2} \right) \left[ 2(1+z)E' - 3E \right] - \frac{(1+z)E}{3 w}\left[2(1+z)(E'^2 + E E'') - 4 EE' \right].
	\end{equation} 
\end{widetext}

From this, we see that using the observed dimensionless Hubble parameter $E(z)$, one can reconstruct the interaction, once the equation of state $w(z)$ of dark energy 
is known. This will be the key equation for the reconstruction of $\tilde{Q}$.

\begin{table*}[t!]
	\caption{{\small The Cosmic Chronometer Hubble parameter $H$ measurements (in units of km s$^{-1}$ Mpc$^{-1}$) and their errors $\sigma_H$ at redshift $z$ obtained from 
	the differential age method (CC).}} 
	\begin{center}
		\resizebox{0.95\textwidth}{!}{\renewcommand{\arraystretch}{1.3} \setlength{\tabcolsep}{18pt}\centering 
			\begin{tabular}{l|c|c|c|c} 
				\hline
				\textbf{Index} & $z$ & $H \pm \sigma_H$ \textbf{(BC03)}& $H \pm \sigma_H$ \textbf{(MaStro11)} & \textbf{References}\\ 
				\hline				
				1 & 0.07 & 69 $\pm$ 19.6 & ... & \cite{cc_101}\\
				2 & 0.1 & 69 $\pm$ 12 & ... & \cite{cc_102}\\
				3 & 0.12 & 68.6 $\pm$ 26.2 & ... & \cite{cc_101}\\
				4 & 0.17 & 83 $\pm$ 8 & ... & \cite{cc_102}\\
				5 & 0.1797 & 75 $\pm$ 4 & 81 $\pm$ 5 & \cite{cc_103}\\
				6 & 0.1993 & 75 $\pm$ 5 & 81 $\pm$ 6 & \cite{cc_103}\\
				7 & 0.2 & 72.9 $\pm$ 29.6 & ...  & \cite{cc_101}\\
				8 & 0.27 & 77 $\pm$ 14 & ...  & \cite{cc_102}\\
				9 & 0.28 & 88.8 $\pm$ 36.6 & ... & \cite{cc_101}\\ 
				10 & 0.3519 & 83 $\pm$ 14 & 88 $\pm$ 16 & \cite{cc_103}\\
				11 & 0.3802 & 83 $\pm$ 13.5 & 89.2 $\pm$ 14.1 & \cite{cc_104}\\
				12 & 0.4 & 95 $\pm$ 17 & ...  & \cite{cc_102}\\
				13 & 0.4004 & 77.0 $\pm$ 10.2 & 82.8 $\pm$ 10.6 & \cite{cc_104}\\
				14 & 0.4247 & 87.1 $\pm$ 11.2 & 93.7 $\pm$ 11.7 & \cite{cc_104}\\
				15 & 0.4497 & 92.8 $\pm$ 12.9 & 99.7 $\pm$ 13.4 & \cite{cc_104}\\
				16 & 0.47 & 89 $\pm$ 34 & ... & \cite{cc_106}\\
				17 & 0.4783 & 80.9 $\pm$ 9.0 &  86.6 $\pm$ 8.7  & \cite{cc_104}\\
				18 & 0.48 & 97 $\pm$ 60 &  ... & \cite{cc_102}\\
				19 & 0.5929 & 104 $\pm$ 13 & 110 $\pm$ 15 & \cite{cc_103}\\
				20 & 0.6797 & 92 $\pm$ 8 & 98 $\pm$ 10 & \cite{cc_103}\\
				21 & 0.7812 & 105 $\pm$ 12 & 88 $\pm$ 11 & \cite{cc_103}\\
				22 & 0.8754 & 125 $\pm$ 17 & 124 $\pm$ 17 & \cite{cc_103}\\
				23 & 0.88 & 90 $\pm$ 40 & ... & \cite{cc_102}\\
				24 & 0.9 & 117 $\pm$ 23 & ... & \cite{cc_102}\\
				25 & 1.037 & 154 $\pm$ 20 & 113 $\pm$ 15 & \cite{cc_103}\\
				26 & 1.3 & 168 $\pm$ 17 & ... & \cite{cc_102}\\
				27 & 1.363 & 160 $\pm$ 33.6 & ... & \cite{cc_105}\\
				28 & 1.43 & 177 $\pm$ 18 & ... & \cite{cc_102}\\
				29 & 1.53 & 140 $\pm$ 14 & ... & \cite{cc_102}\\
				30 & 1.75 & 202 $\pm$ 40 & ... & \cite{cc_102}\\
				31 & 1.965 & 186.5 $\pm$ 50.4 & ... & \cite{cc_105}\\
				\hline
			\end{tabular}
		}
	\end{center}
	\label{tab_CC}
\end{table*}

\section{The Reconstruction}

To reconstruct the interaction using current data sets, we need a model-independent method to reconstruct $E(z)$, and its derivatives $E'(z)$ and $E''(z)$. 
In this work we use the Gaussian Processes (GP)\cite{rw,mackay,william} as a numerical tool for this reconstruction purpose.

\begin{table}[t!]
	\caption{{\small $E(z)$ obtained from the inversion of ${E^{-1}(z)}$ data reported in Table 6 of Ref \cite{mct}. Note the difference in the estimate of $E(z=1.5)$, 
	from the actual quoted value. We pick up the mean value obtained from inversion of quoted ${E^{-1}(z)}$. The inverse covariance matrix has been included in our 
	analysis.}}
	\begin{center}
		\resizebox{0.4\textwidth}{!}{\renewcommand{\arraystretch}{1.3} \setlength{\tabcolsep}{15pt}\centering 
			\begin{tabular}{l|c|c} 
				\hline
				\textbf{Index} & $z$ & $E(z)$ \\ 
				\hline
				1 & 0.07 & 0.997 $\pm$ 0.023 \\
				2 & 0.20 & 1.111 $\pm$ 0.021 \\
				3 & 0.35 & 1.127 $\pm$ 0.037 \\ 
				4 & 0.55 & 1.366 $\pm$ 0.062 \\
				5 & 0.9 & 1.524 $\pm$ 0.121 \\
				6 & 1.5 & 2.924 $\pm$ 0.675 \\
				\hline
			\end{tabular} \label{tabmct}
	} \end{center}
\end{table}

\begin{table}[t!]
	\caption{{\small The Hubble parameter measurements $H(z)$ (in units of km s$^{-1}$ Mpc$^{-1}$) and their errors $\sigma_H$ at redshift $z$ obtained from the radial 
	BAO method (BAO).}}
	\begin{center}
		\resizebox{0.48\textwidth}{!}{\renewcommand{\arraystretch}{1.3} \setlength{\tabcolsep}{9pt}\centering 
			\begin{tabular}{l|c|c|c} 
				\hline
				\textbf{Index} & \textbf{$z$} & \textbf{$H \pm \sigma_H$} & \textbf{References}\\ 
				\hline
				1 & 0.24 & 79.69 $\pm$ 2.99 & \cite{bao_61}\\
				2 & 0.3 & 81.7 $\pm$ 6.22 & \cite{bao_107}\\
				3 & 0.31 & 78.17 $\pm$ 4.74 & \cite{bao_108}\\ 
				4 & 0.34 & 83.8 $\pm$ 3.66 & \cite{bao_61}\\
				5 & 0.35 & 82.7 $\pm$ 8.4 & \cite{bao_77}\\
				6 & 0.36 & 79.93 $\pm$ 3.39 & \cite{bao_108}\\
				7 & 0.38 & 81.5 $\pm$ 1.9 & \cite{bao_10}\\
				8 & 0.40 & 82.04 $\pm$ 2.03 & \cite{bao_108}\\
				9 & 0.43 & 86.45 $\pm$ 3.68 & \cite{bao_61}\\
				10 & 0.44 & 82.6 $\pm$ 7.8 & \cite{bao_73}\\
				11 & 0.44 & 84.81 $\pm$ 1.83 & \cite{bao_108}\\
				12 & 0.48 & 87.79 $\pm$ 2.03 & \cite{bao_108}\\
				13 & 0.51 & 90.4 $\pm$ 1.9 & \cite{bao_10}\\
				14 & 0.52 & 94.35 $\pm$ 2.65 & \cite{bao_108}\\
				15 & 0.56 & 93.33 $\pm$ 2.32 & \cite{bao_108}\\
				16 & 0.57 & 87.6 $\pm$ 7.8 & \cite{bao_2}\\
				17 & 0.57 & 96.8 $\pm$ 3.4 & \cite{bao_80}\\
				18 & 0.59 & 98.48 $\pm$ 3.19 & \cite{bao_108}\\
				19 & 0.6 & 87.9 $\pm$ 6.1 & \cite{bao_73}\\
				20 & 0.61 & 97.3 $\pm$ 2.1 & \cite{bao_10}\\
				21 & 0.64 & 98.82 $\pm$ 2.99 & \cite{bao_108}\\
				22 & 0.73 & 97.3 $\pm$ 7 & \cite{bao_73}\\
				23 & 0.978 & 113.72 $\pm$ 14.63 & \cite{bao_4}\\
				24 & 1.23 & 131.44 $\pm$ 12.42 & \cite{bao_4}\\
				25 & 1.526 & 148.11 $\pm$ 12.71 & \cite{bao_4}\\
				26 & 1.944 & 172.63 $\pm$ 14.79 & \cite{bao_4}\\
				27 & 2.3 & 224 $\pm$ 8	& \cite{bao_1}\\
				28 & 2.33 & 224 $\pm$ 8 & \cite{bao_109}\\
				29 & 2.34 & 222 $\pm$ 7 & \cite{bao_9}\\
				30 & 2.36 & 226 $\pm$ 8 & \cite{bao_110}\\
				31 & 2.4 & 227.8 $\pm$ 5.61 & \cite{bao_3}\\
				\hline
			\end{tabular} \label{tabbao}
	} \end{center}
\end{table}

\subsection{Gaussian Process}

A Gaussian Process (GP) involves an indexed collection of random variables having a Multivariate Normal distribution. GPs can be used to infer a distribution 
over functions directly. The distribution of a GP is the joint distribution of all random variables, which is a distribution over functions within a 
continuous domain. For a given set of Gaussian-distributed observational data, we use GP to reconstruct the most probable underlying continuous 
function describing that data, along with its higher derivatives, and also obtain the associated confidence levels, without limiting to any particular 
parametrization ansatz.\\

Due to its model-independent nature, this method has been widely applied in cosmology. A non-parametric reconstruction using GP has been utilized in 
\cite{holsclaw1, holsclaw2, holsclaw3, 1606.04398[25], 1606.04398[26], 1606.04398[27], 1606.04398[28], 1606.04398[30], yangguocai, wang-meng, wang-meng2, 
purbaq, purba, bilicki, lin, xia, arjona, carlos, jesus_nonpara, nunes, Qlisa}. We refer to the publicly available GP 
website\footnote{\url{http://www.gaussianprocess.org}} for more details of the method.\\

Let us consider a function $f$ formed from a GP. The value of $f$ when evaluated at a redshift point $z$ is a Gaussian random variable with mean $\mu(z)$ and variance 
$\mbox{var}(z)$. The function value at redshift $z$ is not independent of the function value at some other point $\tilde{z}$ (especially when $z$ and $\tilde{z}$ 
are close to each other), but is related by a covariance function $\mbox{cov}(f(z), f(\tilde{z})) = \kappa(z, \tilde{z})$ which correlates the values of different 
$f(z)$ at data points $z$ and $\tilde{z}$ separated by $\vert z-\tilde{z} \vert$ distance units.\\

Thus, the distribution of functions can be described by the following quantities,
\begin{eqnarray}
\mu(z) &=& \mathcal{E}[f(z)], \\
\kappa(z,\tilde{z}) &=&  \mathcal{E}[(f(z)-\mu(z))(f(\tilde{z})-\mu(\tilde{z}))],\\
\mbox{var}(z) &=& \kappa(z,z).
\end{eqnarray}
where $\mathcal{E}$ denotes the expectation.\\ 

The Gaussian process is written as 
\begin{equation}
f(x) \sim \mathcal{GP} (\mu(z), \kappa(z, \tilde{z})),
\end{equation}   where $\mathcal{GP}$ represents a Gaussian Process.\\

The covariance function $\kappa(z, \tilde{z})$ depends on a set of free parameters, called the \textit{hyperparameters}, namely the  characteristic length scale 
$l$ and the signal variance $\sigma_f$. A wide range of possible covariance functions is already present in literature\cite{rw, mackay, william}. As a 
standard choice one may consider the squared exponential covariance, 
\begin{equation} \label{sqexp}
\kappa(z, \tilde{z}) = \sigma_f^2 \exp \left( - \frac{(z-\tilde{z})^2}{2l^2}\right).
\end{equation}
Another possible choice is the Mat\'{e}rn covariance,

\begin{equation}
\begin{split}
\kappa_{\nu=p+\frac{1}{2}}(z,\tilde{z}) = \sigma_f^2 \exp \left( \frac{-\sqrt{2p+1}}{l} \vert z - \tilde{z} \vert \right) \\
\times \frac{p!}{(2p)!} \sum_{i=0}^{p} \frac{(p+i)!}{i!(p-i)!} \left( \frac{2\sqrt{2p+1}}{l} \vert z - \tilde{z} \vert \right)^{p-i}.
\end{split}
\end{equation}

Given a data set $\mathcal{D}$ of $n$ observations, $\mathcal{D} = \left\lbrace(z_i, y_i)\vert_{i = 1, . . . , n}\right\rbrace$, we attempt to 
reconstruct a function $f(z)$ that describes this data. For GPs, any $z_i$ is assigned a random variable $f(z_i)$, and the joint distribution of a finite 
number of these variables $\lbrace f(z_1),\dots,f(z_n)\rbrace$ is itself Gaussian,
\begin{equation}
\mathbf{f} \sim \mathcal{GP}(\boldsymbol\mu, \mathbf{K}) \label{function}.
\end{equation}

For a set of input points $\mathbf{X} = \lbrace z_i \rbrace$, the covariance matrix $\mathbf{K} =\kappa (\mathbf{X}, \mathbf{X}) $ is given by 
$[\kappa(\mathbf{X}, \mathbf{X})]_{ij} = \kappa(z_i, z_j)$. $\boldsymbol\mu = (\mu({z}_1),\cdots,\mu({z}_n))$ and $\mathbf{f} = (f({z}_1),\cdots,f({z}_n))$. 
Excluding observational data, we use the covariance matrix $\mathbf{K}$ to generate a Gaussian vector $\mathbf{f}^{*}$ of function values at $\mathbf{X}^{*}$ 
with $f^{*}_i = f(z^{*}_i)$ such that,

\begin{equation}\label{prior}
\mathbf{f}^{*} \sim \mathcal{GP} \left( \boldsymbol {\mu}^{*} , \mathbf{K}^{**} \right),
\end{equation} 
where $\boldsymbol{\mu}^{*}$ is the a priori assumed mean of $\mathbf{f}^{*}$, and $\mathbf{K}^{**} = \kappa(\mathbf{X}^{*}, \mathbf{X}^{*})$.\\

Observational data $(z_i, y_i)$ can also be described by GPs, on assuming that their errors are Gaussian. The actual observations are assumed to be scattered 
around the underlying function, i.e. $y_i = f(x_i)+\epsilon_i$, where Gaussian noise $\epsilon_i$ with variance $\sigma_i^2$ is assumed. This variance needs 
to be added to the covariance matrix,
\begin{equation}\label{gpdata}
\mathbf{y} \sim \mathcal{GP} \left( \boldsymbol{\mu}, \mathbf{K} + \mathbf{C} \right),
\end{equation} 

where $\mathbf{C}$ is the covariance matrix of the data. For uncorrelated data, we use $\mathbf{C} = \sigma_i^2 \mathbf{I}$. The above two GPs, 
\eqref{prior} for $\mathbf{f}^*$ and \eqref{gpdata} for $\mathbf{y}$ can be combined in the joint distribution,
\begin{equation}
\begin{bmatrix}
\mathbf{y} \\
\mathbf{f}^*
\end{bmatrix} \sim \mathcal{GP} \left( \begin{bmatrix}
\boldsymbol{\mu} \\
\boldsymbol{\mu}^*
\end{bmatrix} \begin{bmatrix}
\mathbf{K} + \mathbf{C} & \mathbf{K}^* \\
{\mathbf{K}^{*}}^{\mbox{\tiny T}} & \mathbf{K}^{**} 	\end{bmatrix}\right),
\end{equation} 

where $\mathbf{K}^* = \kappa(\mathbf{X},\mathbf{X}^*)$ and ${\mathbf{K}^{*}}^{\mbox{\tiny T}} = \kappa(\mathbf{X}^*,\mathbf{X})$. While $\mathbf{y}$ is known 
from observations, we want to reconstruct $\mathbf{f}^*$. \\

Using standard rules for conditioning Gaussians, the predictive distribution is given by,
\begin{equation}\label{posterior}
\mathbf{f}^* \vert \mathbf{X}^*, \mathbf{X}, \mathbf{y} \sim \mathcal{GP}	\left(\mathbf{\boldsymbol{\overline{\mu}}}^*, \boldsymbol{\Sigma}^* \right),
\end{equation} 
where

\begin{equation}
\mathbf{\boldsymbol{\overline{\mu}}}^* = \boldsymbol{\mu}^* + {\mathbf{K}^{*}}^{\mbox{\tiny T}} [\mathbf{K} + \mathbf{C}]^{-1} (\mathbf{y} - \boldsymbol{\mu}),
\end{equation} 
and

\begin{equation}
\boldsymbol{\Sigma}^* =  \mathbf{K}^{**} - {\mathbf{K}^{*}}^{\mbox{\tiny T}} [\mathbf{K}+\mathbf{C}]^{-1} \mathbf{K}^*	
\end{equation} 
are the mean and covariance of $\mathbf{f}^*$ respectively.
Equation \eqref{posterior} is the posterior distribution of the function given the data \eqref{gpdata} and the prior \eqref{prior}.\\

Although equation \eqref{posterior} covers noise in training data, it is still a distribution over noise-free predictions $\mathbf{f}^*$. To include noise 
$\boldsymbol{\epsilon}$ into predictions $\mathbf{y}^*$ we add $\sigma_i^2$ to the diagonal of $\boldsymbol{\Sigma}^*$, i.e.,

\begin{equation}\label{posterior_noise}
\mathbf{y}^* \vert \mathbf{X}^*, \mathbf{X}, \mathbf{y} \sim \mathcal{GP}	\left(\mathbf{\boldsymbol{\overline{\mu}}}^*, \boldsymbol{\Sigma}^*+\mathbf{C} \right).
\end{equation} 

To apply the above equations for reconstructing a function, we need to estimate the hyperparameters $\sigma_f$ and $l$. They can be trained by maximizing 
the marginal likelihood. Note that the marginal likelihood depends only on the locations $\mathbf{X}$ of the observations, but not on the points $\mathbf{X}^*$, 
where we want to reconstruct the function. For a Gaussian prior $\mathbf{f}\vert \mathbf{X}, \sigma_f , l \sim \mathcal{GP} (\boldsymbol{\mu}, \mathbf{K})$ 
and with $\mathbf{y} \vert \mathbf{f} \sim \mathcal{GP} (\mathbf{f}, \mathbf{C})$, the log marginal likelihood is given by,
\begin{equation}\label{lnlike}
\ln \mathcal{L} = -	\frac{1}{2}(\mathbf{y}-\boldsymbol{\mu})^{\mbox{\tiny T}} [\mathbf{K} + \mathbf{C}]^{-1} ~(\mathbf{y}-\boldsymbol{\mu}) -\frac{1}{2}\ln \vert \mathbf{K} + \mathbf{C} \vert - \frac{n}{2}\ln 2\mathrm{\pi}.
\end{equation}
The hyperparameters $\sigma_f$ and $l$ are optimized by maximizing equation \eqref{lnlike}.\\

\begin{table*}[t!] 
	\caption{{\small Table showing the reconstructed value of $H_0$ using samples from Set A and B.}}
	\begin{center}
		\resizebox{0.999\textwidth}{!}{\renewcommand{\arraystretch}{1.5} \setlength{\tabcolsep}{27pt} \centering  
			\begin{tabular}{|l|c|c|c|c|} 
				\hline
				$\kappa(z,\tilde{z}$)& \textbf{A 1} & \textbf{A 2} & \textbf{B 1} & \textbf{B 2} \\
				\hline
				Sq. Exp & $67.36\pm4.77  $ & $ 72.13\pm4.85$ & $65.19\pm2.63$ & $67.66\pm2.79$\\ 
				\hline				
				Mat\'{e}rn 9/2 & $68.47\pm5.08$ & $72.76\pm5.00$ & $65.15\pm2.72$ & $67.57\pm2.90$\\ 
				\hline
			\end{tabular}
		}
	\end{center}
	\label{Hz_res}
\end{table*}

This approach also provides a robust way to estimate derivatives of the function. While the covariance between the observational points stays the same, 
one also needs a covariance between the function and its derivative and another between the derivatives for the reconstruction. These covariances can be 
obtained by differentiating the original covariance function $\kappa(x,\tilde{x})$. \\

In the present work, the observational Hubble data from Cosmic Chronometers and Baryon Acoustic Oscillations, along with the 
reduced Hubble data from the Pantheon Supernova compilation of CANDELS and CLASH Multy-Cycle Treasury programs, forms the function 
space. The target function is $E(z)$, and its derivatives $E'(z)$ and $E''(z)$, which are to be reconstructed using the GP method.\\
 
We have used both the squared exponential and the Mat\'{e}rn $(\nu = \frac{9}{2},~ p = 4)$ covariance functions. The former has the advantage of being indefinitely 
differentiable. The latter,  Mat\'{e}rn $(\nu)$ leads to reliable and stable results when $\nu > n$ where derivatives up to $n$th order is required. In the present work, we need 
to use up to second order derivative, so $\nu$ is easily greater that $n$. For a comprehensive comparison of various covariance functions, we refer to the work of Seikel and 
Clarkson\cite{seikel13}. In contrast to actual parameters, GP does not specify the form of the reconstructed function. It rather characterizes the typical changes in the 
function. The publicly available \texttt{GaPP}\footnote{\url{https://github.com/carlosandrepaes/GaPP}} (Gaussian Processes in Python)  code  by Seikel, Clarkson and 
Smith \cite{1606.04398[25]} has been employed in this work.

\subsection{Observational Datasets}

The background data viz. the Cosmic Chronometer (CC) Hubble data, the Pantheon Supernova compilation of CANDELS and CLASH Multy-Cycle Treasury (MCT) programs obtained by the HST, 
and the Baryon Acoustic Oscillation (BAO) Hubble data are utilized for reconstructing the cosmic interaction as a function of redshift. A brief summary of the datasets is given 
below.

\subsubsection{CC Data}

The Hubble parameter $H(z)$ are estimated by calculating the differential ages of galaxies \cite{cc_101,cc_102,cc_103,cc_104,cc_106,cc_105}, usually called cosmic chronometer (CC),
as
\begin{equation} \label{ccH}
H(z) = -\frac{1}{1+z} \frac{dz}{dt}.
\end{equation}
The CC data are independent of the Cepheid distance scale and do not assume any background cosmological model. However, they are subject to other sources of systematic 
uncertainties, such as the modelling of stellar ages carried out through the stellar population synthesis techniques (SPS). Given a pair of ensembles of passively-evolving 
galaxies at two different redshift it is possible to infer $\frac{\Delta z}{\Delta t}$ from observations under the assumption of a concrete SPS model \cite{sps1, sps2, sps3}. 
Thus $H(z)$ can directly be computed using equation \eqref{ccH}, the quantity we are interested in. Therefore, the CC measurements allow us to obtain direct information about 
the Hubble function at different $z$, contrary to other probes which do not directly measure $H(z)$, but integrated quantities such as luminosity distances. In the present work, 
we use both the BC03 and MaStro11 SPS compilation of CC measurements shown in Table \ref{tab_CC} including almost all $H(z)$ data reported in various surveys so far. The sources 
of these datasets are quoted in the table.

\begin{figure*} [t!]
	\begin{center}
		\includegraphics[angle=0, width=\textwidth]{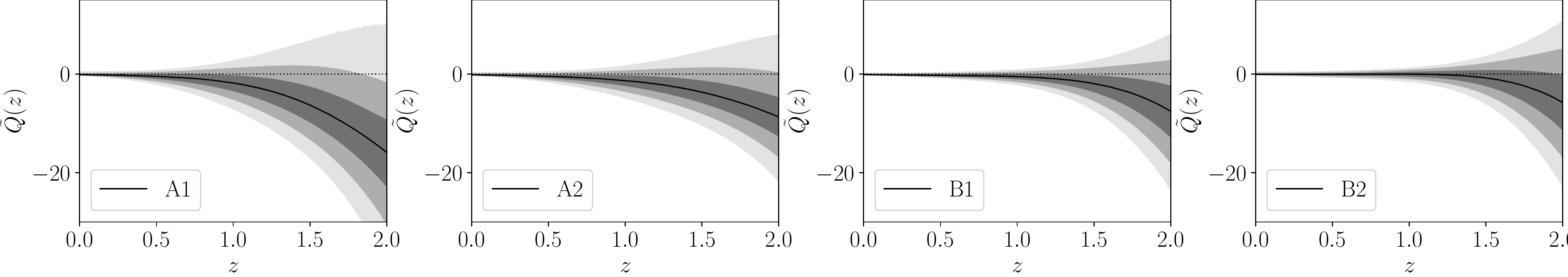}\\
		\includegraphics[angle=0, width=\textwidth]{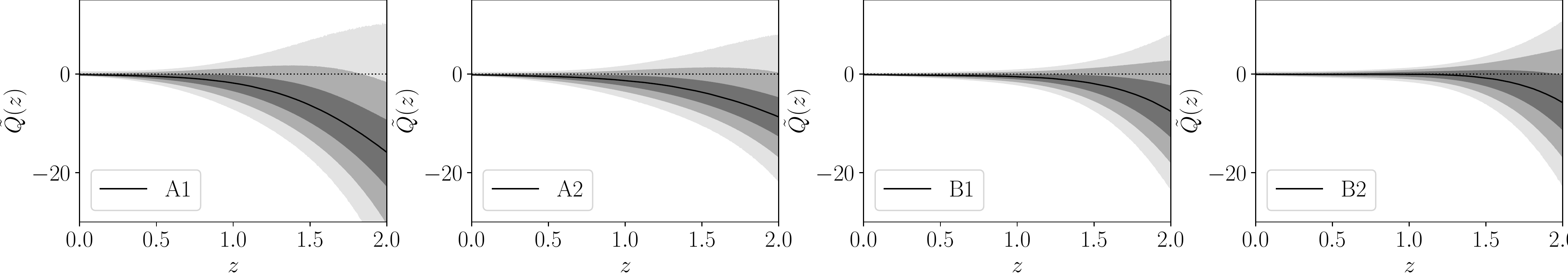}\\
		\includegraphics[angle=0, width=\textwidth]{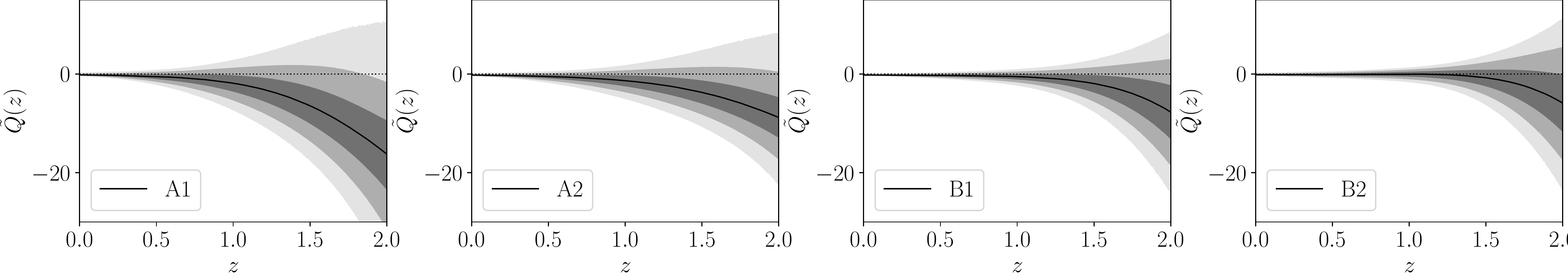}
	\end{center}
	\caption{{\small Plots for the reconstructed interaction term $\tilde{Q}$ from the dataset samples of Set A and B using a Squared Exponential Covariance function 
	considering the decaying Dark Energy EoS given by $w = -1$ (top row), the $w$CDM model with DE EoS given by $w = -1.006 \pm 0.045$\cite{wcdm} (middle row), and the CPL 
	parametrization of Dark Energy with DE EoS given by $w(z) = w_0 + w_a (\frac{z}{1+z})$, $w_0 = -1.046^{+0.179}_{-0.170}$ and $w_a = 0.14^{+0.60}_{-0.76}$\cite{cpl} 
	(bottom row). The black solid curve shows the best fit values and the shaded regions correspond to the 1$\sigma$, 2$\sigma$ and 3$\sigma$ uncertainty.}}
	\label{Q_sqexp}
\end{figure*}

\begin{figure*} [t!]
	\begin{center}
		\includegraphics[angle=0, width=\textwidth]{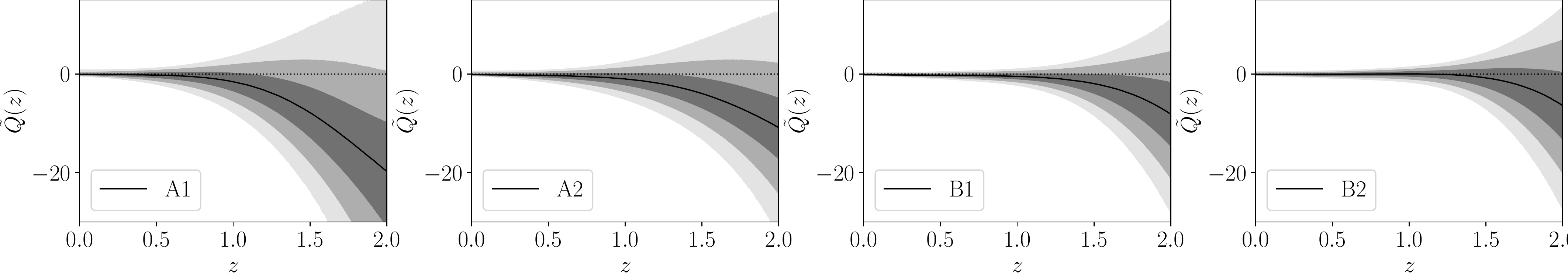}\\
		\includegraphics[angle=0, width=\textwidth]{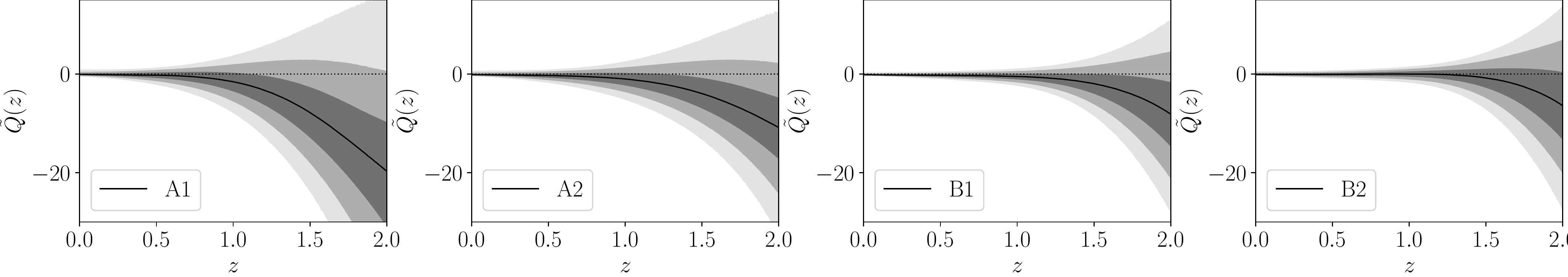}\\
		\includegraphics[angle=0, width=\textwidth]{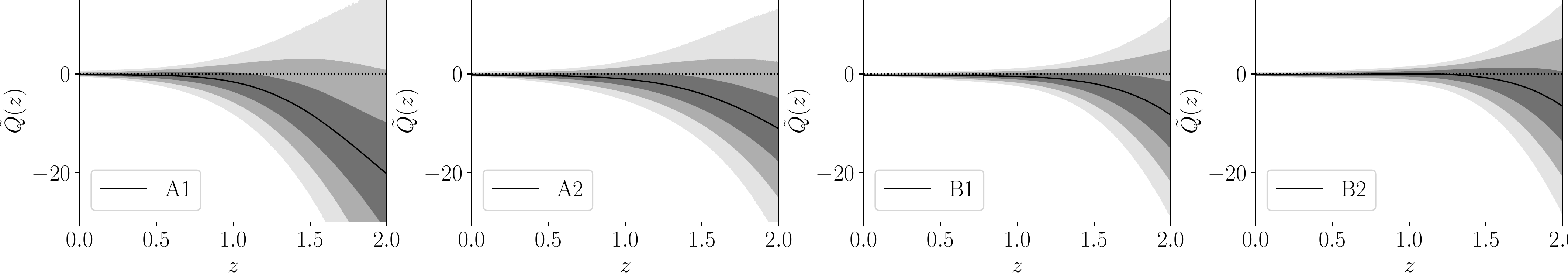}
	\end{center}
	\caption{{\small Plots for the reconstructed interaction term $\tilde{Q}$ from the dataset samples of Set A and B using a Mat\'{e}rn 9/2 Covariance function considering 
	the decaying Dark Energy EoS given by $w = -1$ (top row), the $w$CDM model with DE EoS given by $w = -1.006 \pm 0.045$\cite{wcdm} (middle row), and the CPL parametrization 
	of Dark Energy with EoS given by $w(z) = w_0 + w_a (\frac{z}{1+z})$, $w_0 = -1.046^{+0.179}_{-0.170}$ and $w_a = 0.14^{+0.60}_{-0.76}$\cite{cpl} (bottom row). The black 
	solid curve shows the best fit values and the shaded regions correspond to the 1$\sigma$, 2$\sigma$ and 3$\sigma$ uncertainty.}}
	\label{Q_mat92}
\end{figure*}

\begin{figure*} [t!]
	\begin{center}
		\includegraphics[angle=0, width=\textwidth]{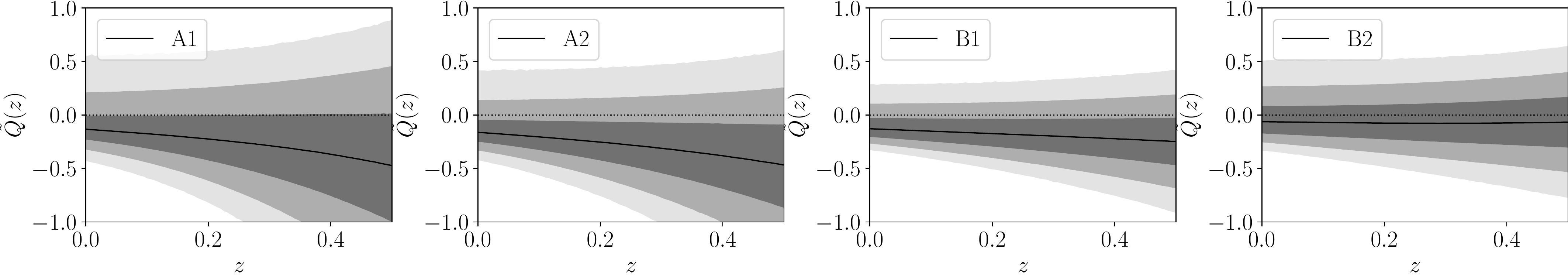}\\
		\includegraphics[angle=0, width=\textwidth]{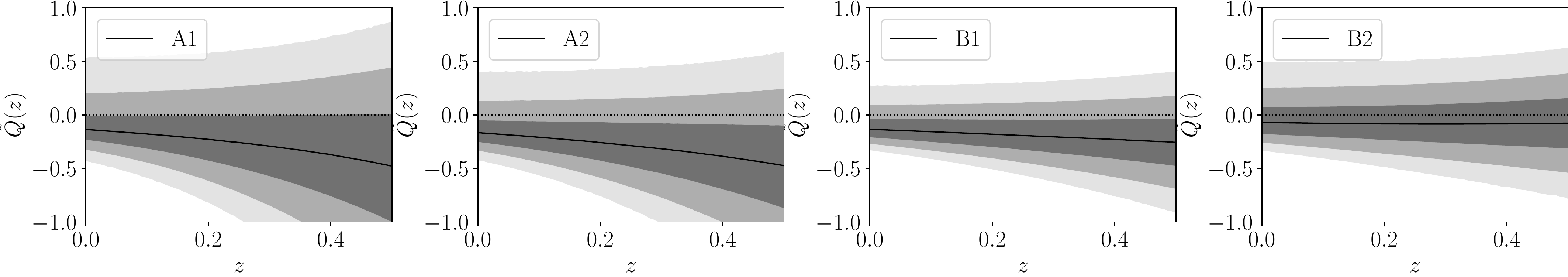}\\
		\includegraphics[angle=0, width=\textwidth]{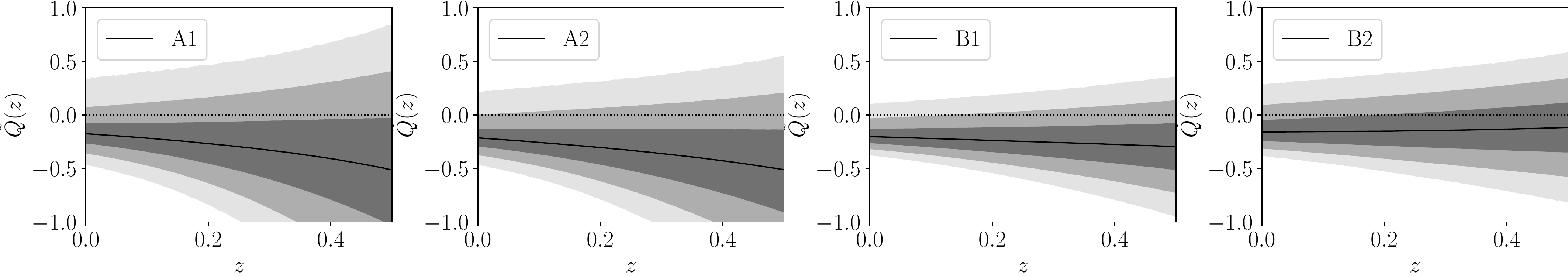}
	\end{center}
	\caption{{\small Plots for the reconstructed $\tilde{Q}$ function in the low redshift range $0<z<0.5$ from the dataset samples of Set A and B using a Squared 
	Exponential Covariance function considering the decaying Dark Energy EoS given by $w = -1$ (top row), the $w$CDM model with DE EoS given by 
	$w = -1.006 \pm 0.045$\cite{wcdm} (middle row), and the CPL parametrization of Dark Energy with DE EoS given by 
	$w(z) = w_0 + w_a (\frac{z}{1+z})$, $w_0 = -1.046^{+0.179}_{-0.170}$ and $w_a = 0.14^{+0.60}_{-0.76}$\cite{cpl} (bottom row). The black solid curve shows the best 
	fit values and the shaded regions correspond to the 1$\sigma$, 2$\sigma$ and 3$\sigma$ uncertainty.}}
	\label{Qscaled_sqexp}
\end{figure*}

\begin{figure*} [t!]
	\begin{center}
		\includegraphics[angle=0, width=\textwidth]{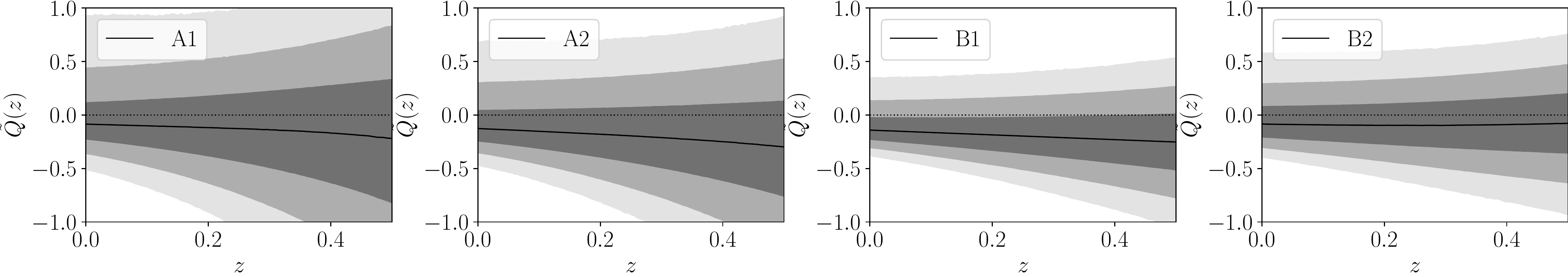}\\
		\includegraphics[angle=0, width=\textwidth]{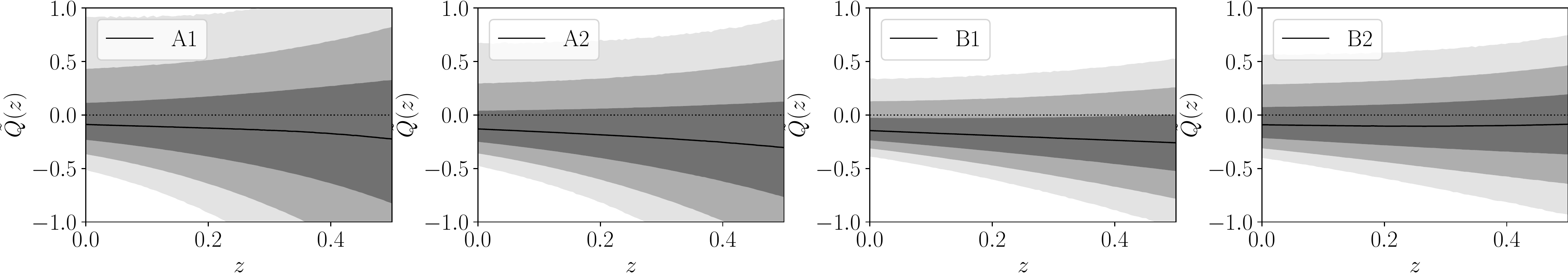}\\
		\includegraphics[angle=0, width=\textwidth]{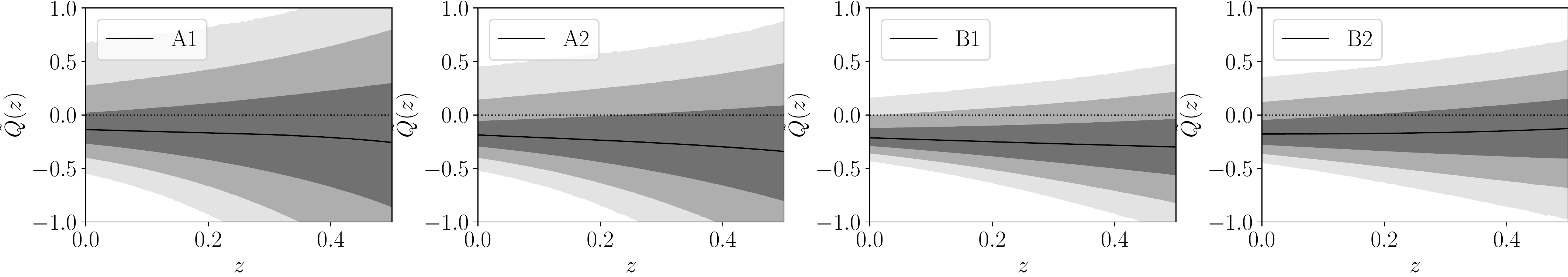}
	\end{center}
	\caption{{\small Plots for the reconstructed $\tilde{Q}$ function in the low redshift range $0<z<0.5$ from the dataset samples of Set A and B using a 
	Mat\'{e}rn 9/2 Covariance function considering the decaying Dark Energy EoS given by $w = -1$ (top row), the $w$CDM model with DE EoS given by 
	$w = -1.006 \pm 0.045$\cite{wcdm} (middle row), and the CPL parametrization of Dark Energy with EoS given by 
	$w(z) = w_0 + w_a (\frac{z}{1+z})$, $w_0 = -1.046^{+0.179}_{-0.170}$ and $w_a = 0.14^{+0.60}_{-0.76}$\cite{cpl} (bottom row). The black solid curve shows the best 
	fit values and the shaded regions correspond to the 1$\sigma$, 2$\sigma$ and 3$\sigma$ uncertainty.}}
	\label{Qscaled_mat92}
\end{figure*}

\begin{table*}[t!] 
	\caption{{\small Table showing the reconstructed value of $\tilde{Q}(z=0)$ using samples from Set A and B for the decaying Dark Energy EoS given by $w = -1$.}}
	\begin{center}
		\resizebox{0.999\textwidth}{!}{\renewcommand{\arraystretch}{1.5} \setlength{\tabcolsep}{5pt} \centering  
			\begin{tabular}{|c|c|c|c|c|} 
				\hline
				$\kappa(z,\tilde{z})$& \textbf{A 1} & \textbf{A 2} & \textbf{B 1} & \textbf{B 2} \\
				\hline
				Sq. Exp & $-0.133^{+0.126~+0.343~+0.686}_{-0.096~-0.190~-0.296}$ & $-0.162^{+0.119~+0.301~+0.581}_{-0.088~-0.170~-0.261}$ & $-0.129^{+0.102~+0.234~+0.417}_{-0.076~-0.138~-0.198}$ & $-0.063^{+0.146~+0.331~+0.571}_{-0.109~-0.193~-0.269}$ \\ 
				\hline				
				Mat\'{e}rn 9/2 & $-0.085^{+0.204~+0.529~+1.016}_{-0.145~-0.280~-0.429}$ & $-0.127^{+0.174~+0.433~+0.812}_{-0.122~-0.230~-0.347}$ & $-0.141^{+0.121~+0.280~+0.492}_{-0.092~-0.168~-0.243}$ & $-0.085^{+0.168~+0.381~+0.671}_{-0.126~-0.224~-0.316}$ \\ 
				\hline
			\end{tabular}
		}
	\end{center}
	\label{Q_lcdm_res}
\end{table*}

\begin{table*}[t!] 
	\caption{{\small Table showing the reconstructed value of $\tilde{Q}(z=0)$ using samples from Set A and B for the $w$CDM model with EoS given by 
	$w = -1.006 \pm 0.045$\cite{wcdm}.}}
	\begin{center}
		\resizebox{0.999\textwidth}{!}{\renewcommand{\arraystretch}{1.5} \setlength{\tabcolsep}{5pt} \centering  
			\begin{tabular}{|c|c|c|c|c|} 
				\hline
				$\kappa(z,\tilde{z})$& \textbf{A 1} & \textbf{A 2} & \textbf{B 1} & \textbf{B 2} \\
				\hline
				Sq. Exp & $-0.135^{+0.124~+0.335~+0.670}_{-0.096~-0.190~-0.296}$ & $-0.165^{+0.116~+0.296~+0.567}_{-0.086~-0.168~-0.257}$ & $-0.133^{+0.099~+0.228~+0.409}_{-0.075~-0.136~-0.197}$ & $-0.069^{+0.143~+0.324~+0.564}_{-0.107~-0.189~-0.266}$ \\ 
				\hline				
				Mat\'{e}rn 9/2 & $-0.088^{+0.202~+0.518~+1.008}_{-0.143~-0.278~-0.425}$ & $-0.130^{+0.170~+0.425~+0.804}_{-0.120~-0.229~-0.345}$ & $-0.146^{+0.118~+0.273~+0.488}_{-0.091~-0.167~-0.244}$ & $-0.091^{+0.165~+0.376~+0.653}_{-0.124~-0.219~-0.311}$ \\ 
				\hline
			\end{tabular}
		}
	\end{center}
	\label{Q_wcdm_res}
\end{table*}

\begin{table*}[t!] 
	\caption{{\small Table showing the reconstructed value of $\tilde{Q}(z=0)$ using samples from Set A and B for the CPL parametrization of Dark Energy with EoS given 
	by $w(z) = w_0 + w_a (\frac{z}{1+z})$, $w_0 = -1.046^{+0.179}_{-0.170}$ and $w_a = 0.14^{+0.60}_{-0.76}$\cite{cpl}.}}
	\begin{center}
		\resizebox{0.999\textwidth}{!}{\renewcommand{\arraystretch}{1.5} \setlength{\tabcolsep}{5pt} \centering  
			\begin{tabular}{|c|c|c|c|c|} 
				\hline
				$\kappa(z,\tilde{z})$& \textbf{A 1} & \textbf{A 2} & \textbf{B 1} & \textbf{B 2} \\
				\hline
				Sq. Exp & $-0.175^{+0.097~+0.249~+0.520}_{-0.091~-0.185~-0.293}$ & $-0.214^{+0.087~+0.217~+0.429}_{-0.080~-0.162~-0.253}$ & $-0.202^{+0.073~+0.169~+0.306}_{-0.061~-0.117~-0.177}$ & $-0.159^{+0.111~+0.253~+0.448}_{-0.085~-0.155~-0.255}$ \\ 
				\hline				
				Mat\'{e}rn 9/2 & $-0.136^{+0.157~+0.410~+0.814}_{-0.132~-0.263~-0.412}$ & $-0.187^{+0.131~+0.330~+0.638}_{-0.108~-0.214~-0.331}$ & $-0.213^{+0.091~+0.209~+0.374}_{-0.076~-0.146~-0.221}$ & $-0.178^{+0.131~+0.299~+0.530}_{-0.100~-0.186~-0.272}$ \\ 
				\hline
			\end{tabular}
		}
	\end{center}
	\label{Q_cpl_res}
\end{table*}

\subsubsection{Pantheon+MCT Data}

We make use of the Hubble rate data points, i.e. $E(z_i )$ = $H(z_i )/H_0$ provided in \cite{mct} for six different redshift in the range $z \in [0.07, 1.5]$. They compress 
information very effectively about the 1048 SN-Ia data at $z < 1.5$ that is a part of the Pantheon compilation (which includes 740 SN-Ia of the joint light-curve analysis (JLA) 
sample), and the 15 SN-Ia at $z > 1$ of the CANDELS and CLASH Multy-Cycle Treasury (MCT) programs obtained by the HST, 9 of which are at $1.5 < z < 2.3$. Riess \textit{et al.}\cite{mct} 
converted the raw SN-Ia measurements into data on $E(z)$ by parametrizing $E^{-1} (z)$. The corresponding values of $E^{-1} (z_i )$ are Gaussian in a 
very good approximation as shown in the work of Riess \textit{et al.}\cite{mct} which also contains the corresponding correlation matrix. Here, we have adopted the mean value 
obtained from inversion of quoted ${E^{-1}(z)}$ and the inverse covariance matrix.

\subsubsection{BAO data}

An alternative compilation of the $H(z)$ data can be deduced from  the radial BAO peaks in the galaxy power spectrum, or from the BAO peak using the Ly-$\alpha$ forest of QSOs, 
which are based on the clustering of galaxies or quasars \cite{bao_61,bao_10,bao_107,bao_108,bao_109,bao_110,bao_1,bao_2,bao_4,bao_73,bao_77,bao_80,bao_9,bao_3}. Table \ref{tabbao}
includes almost all data reported in various surveys so far. One may find that some of the $H(z)$ data points from clustering measurements are correlated since they either belong 
to the same analysis or there is an overlap between the galaxy samples. We take into account all covariances among the respective data points in our analysis.

\subsection{Methodology}

We consider three choices for the dark energy equation of state parameter $w = \frac{p_D}{{\rho}_D}$. First we consider the decaying vacuum energy case, followed by 
the $w$CDM model and finally the CPL parametrization, 
\begin{eqnarray}
\Lambda\mbox{CDM}	: w(z) &=& -1 \\
w\mbox{CDM}	:  w(z) &=& w  \\
\mbox{CPL}	 :  w(z) &=& w_0 + w_a \left(\frac{z}{1+z}\right) 
\end{eqnarray}
For the dark energy EoS, considering the $w$CDM model, we take the best-fit value of $w = -1.006 \pm 0.045$ from the Planck 2015 \cite{wcdm}, and for the 
CPL parametrization $w(z) = w_0 + w_a (\frac{z}{1+z})$, we take $w_0 = -1.046^{+0.179}_{-0.170}$ and $w_a = 0.14^{+0.60}_{-0.76}$ respectively from HST Cluster Supernova Survey 
2011 \cite{cpl}.\\

We attempt to reconstruct $\tilde{Q}$ directly through the Gaussian Process for the following combination of datasets:
\begin{itemize}
	\item \textbf{Set A}
\begin{enumerate}
	\item CC(BC03)+Pantheon+MCT
	\item CC(MaStro)+Pantheon+MCT
\end{enumerate}
	\item \textbf{Set B}
\begin{enumerate}
\item CC(BC03)+Pantheon+MCT+BAO
\item CC(MaStro)+Pantheon+MCT+BAO
\end{enumerate}	
\end{itemize}

We start with constraining the Hubble parameter in the present epoch ($H_0$).  With the Hubble data, 
we utilize the GP method to reconstruct $H(z)$. The value of $H_0$ obtained is shown in Table \ref{Hz_res}. Further, we normalize the reconstructed dataset to obtain the 
dimensionless or reduced Hubble parameter $E(z) = \frac{H(z)}{H_0}$. Considering the error of Hubble constant $\sigma_H$, we can calculate the uncertainty associated with 
$E$, i.e. $\sigma_{E}$ as,
\begin{equation} \label{sigh_recon}
{\sigma_E}^2 = \frac{{\sigma_H}^2}{ {H_0}^2} + \frac{H^2}{{H_0}^4}{\sigma_{H_0}}^2,
\end{equation} 
where $\sigma_{H_0}$ is the error associated with $H_0$.\\

We normalize the CC and BAO Hubble data with the reconstructed $H_0$ to obtain $E (=\frac{H}{H_{0}})$ with $E(z=0) = 1$. 
These are now combined with $E$ data obtained from the Pantheon+MCT compilation. The error uncertainties and the covariance matrix associated with individual 
data sets have been combined and taken into account. Assuming that the $E$ data, obey a Gaussian distribution with a mean and variance, the posterior 
distribution of reconstructed function $E(z)$ and its derivatives can be expressed as a joint Gaussian distribution of different datasets. Thus, given a 
set of observational data we have used the GP to construct the most probable underlying continuous function $E(z)$ describing the data, along with its 
derivatives, and have also obtained the associated confidence levels. Using the reconstructed values of $E(z)$, $E'(z)$ and $E''(z)$ in Eq. \eqref{Q_h}, 
the interaction $\tilde{Q}(z)$ is reconstructed.

\subsection{Results}

\begin{figure*}[t!]
	\begin{center}
		\includegraphics[angle=0, width=0.25\textwidth]{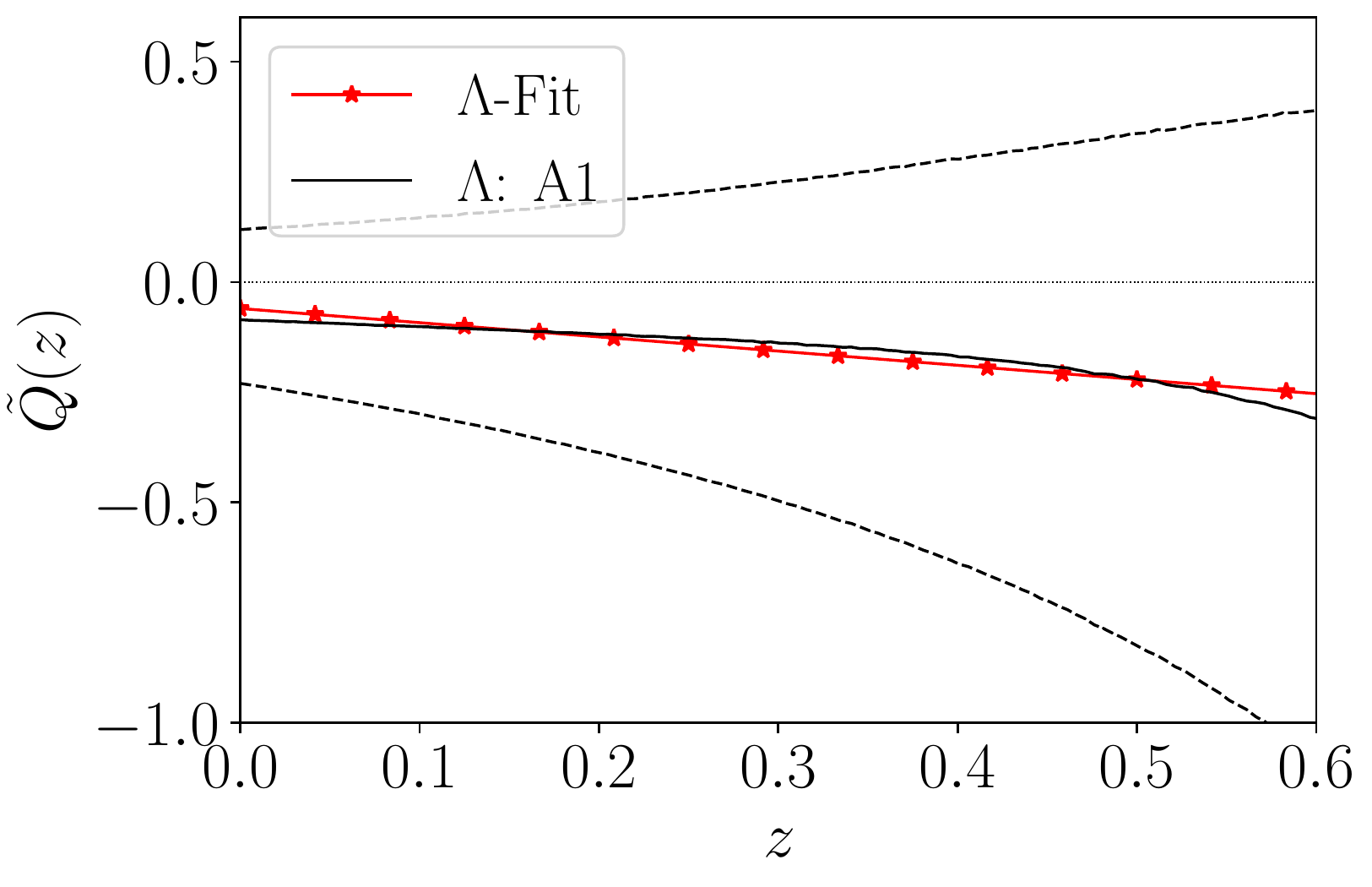}
		\includegraphics[angle=0, width=0.25\textwidth]{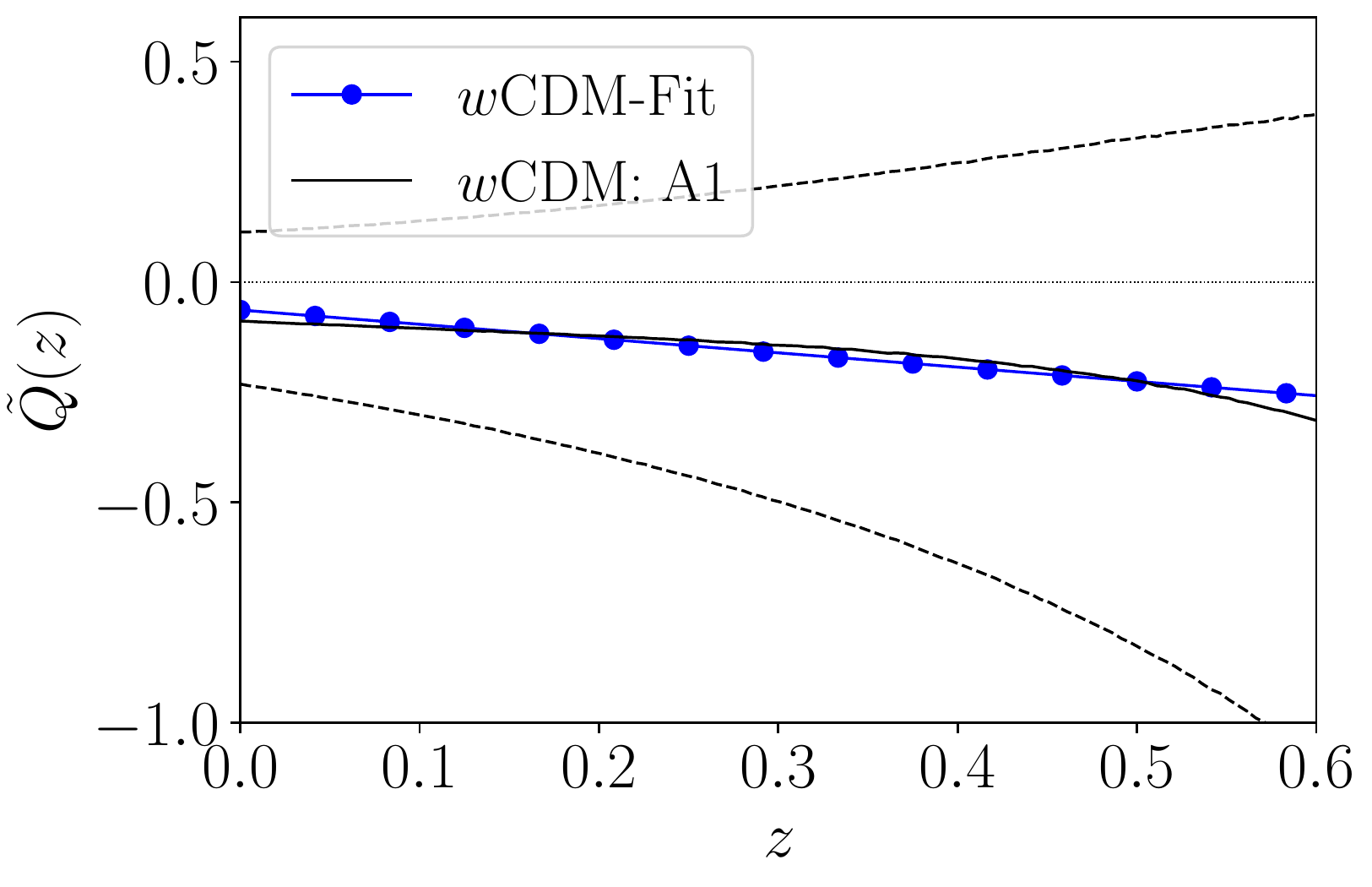}
		\includegraphics[angle=0, width=0.25\textwidth]{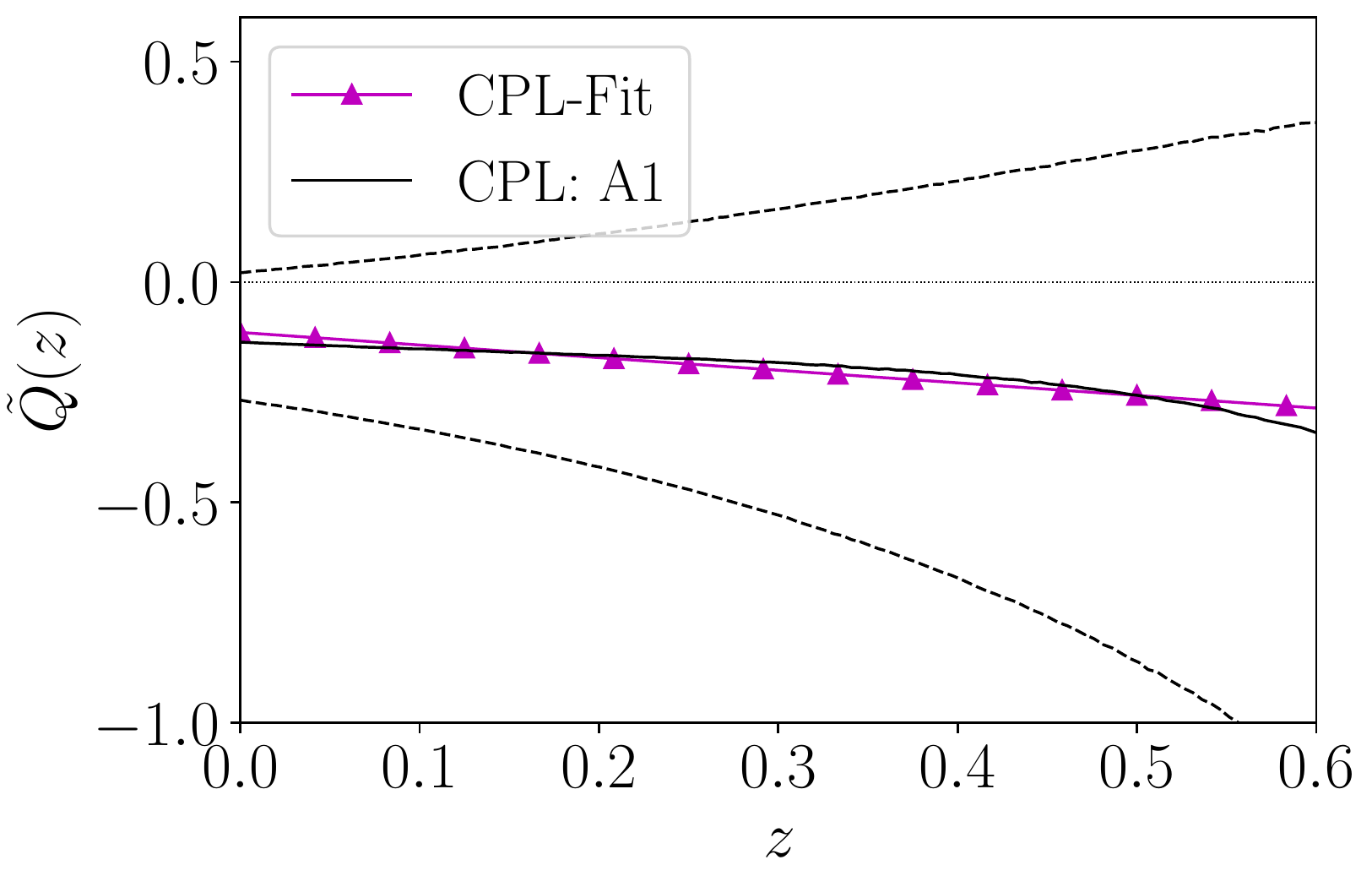}\\
		\includegraphics[angle=0, width=0.25\textwidth]{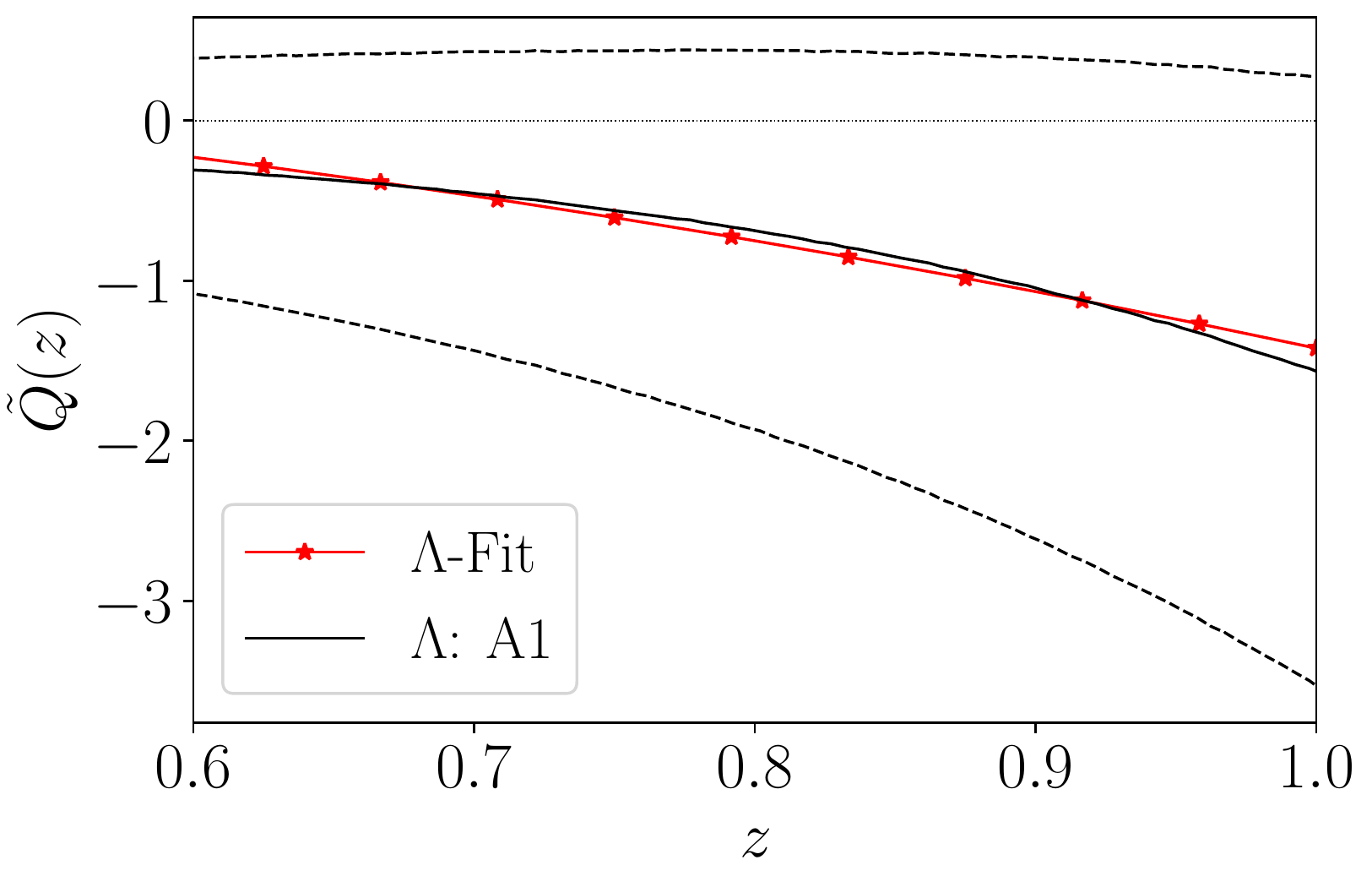}
		\includegraphics[angle=0, width=0.25\textwidth]{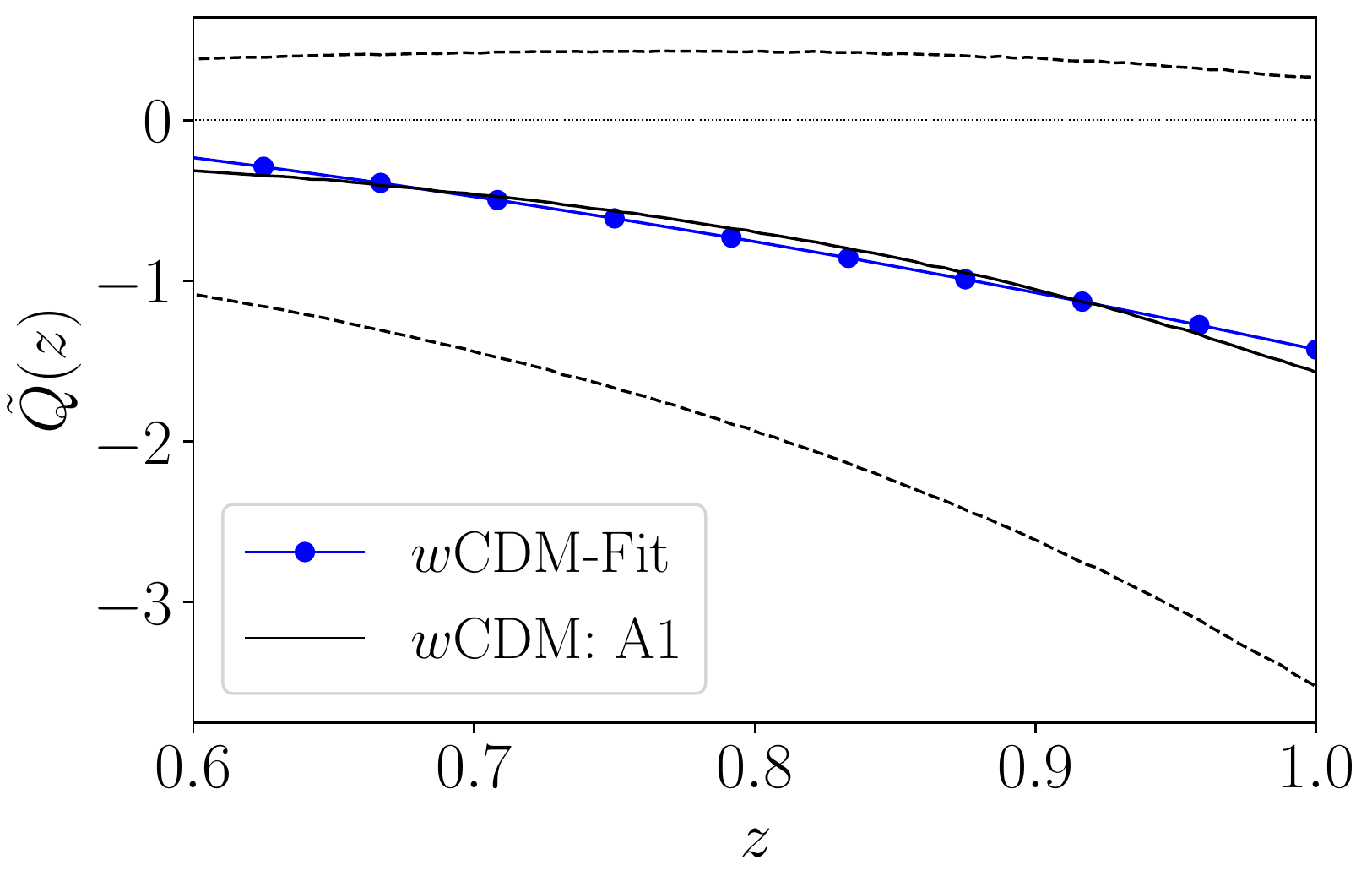}
		\includegraphics[angle=0, width=0.25\textwidth]{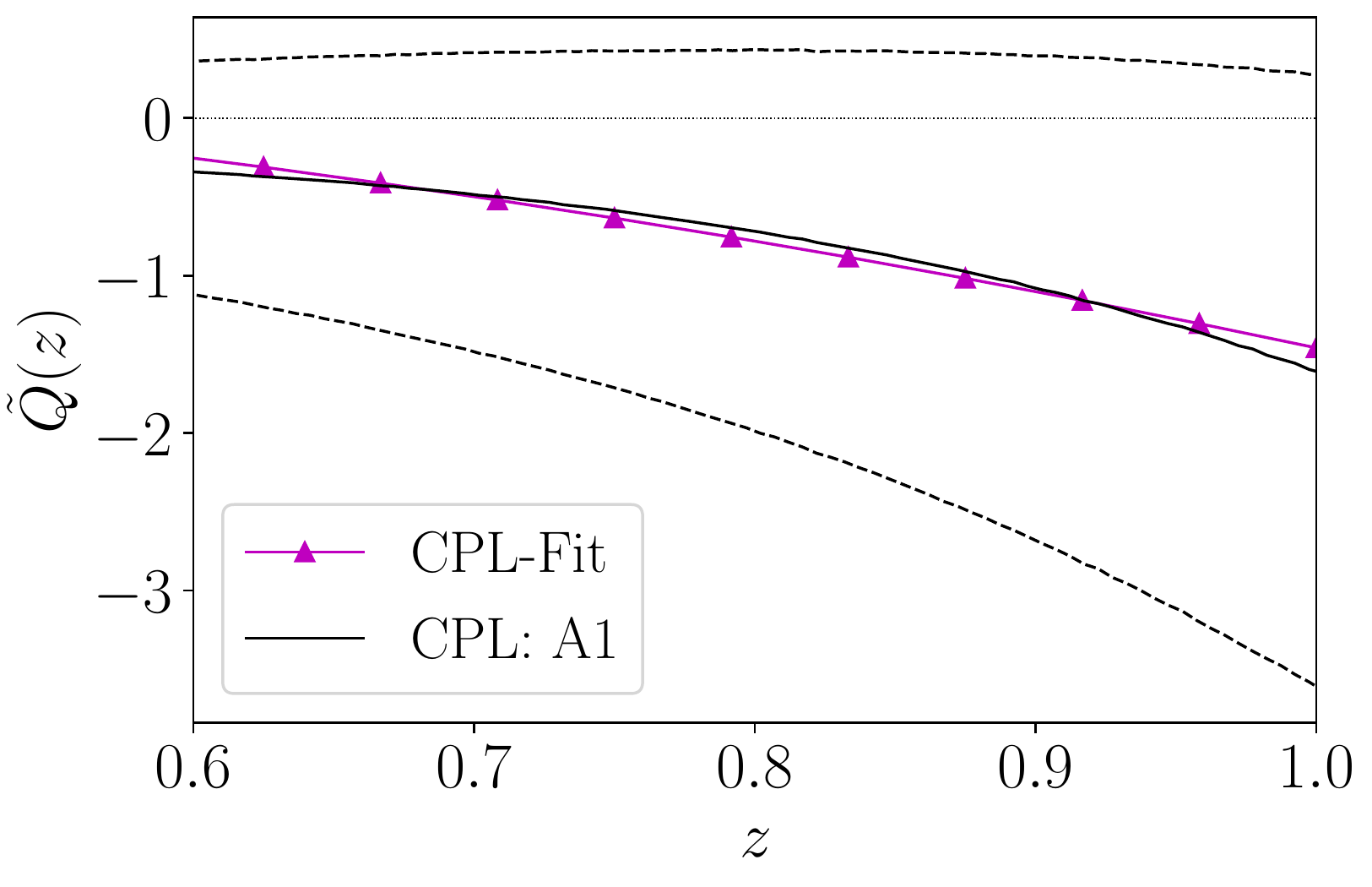}
	\end{center}
	\caption{{\small Plots showing a comparison between the reconstructed interaction $\tilde{Q}(z)$ and the estimated $\tilde{Q}_{\mbox{\tiny fit}}(z)$ using 
			the combined dataset A1, for EoS given by $w=-1$(left), $w$CDM model (middle) and the CPL parametrization (right). The black solid line is 
			the reconstructed function. The line with marker represents the best fit result from $\chi^2$-minimization. The 1$\sigma$ C.L.s are shown in dashed 
			lines.}}
	\label{Qfit_plot1}
\end{figure*}

\begin{figure*}
	\begin{center}
		\includegraphics[angle=0, width=0.25\textwidth]{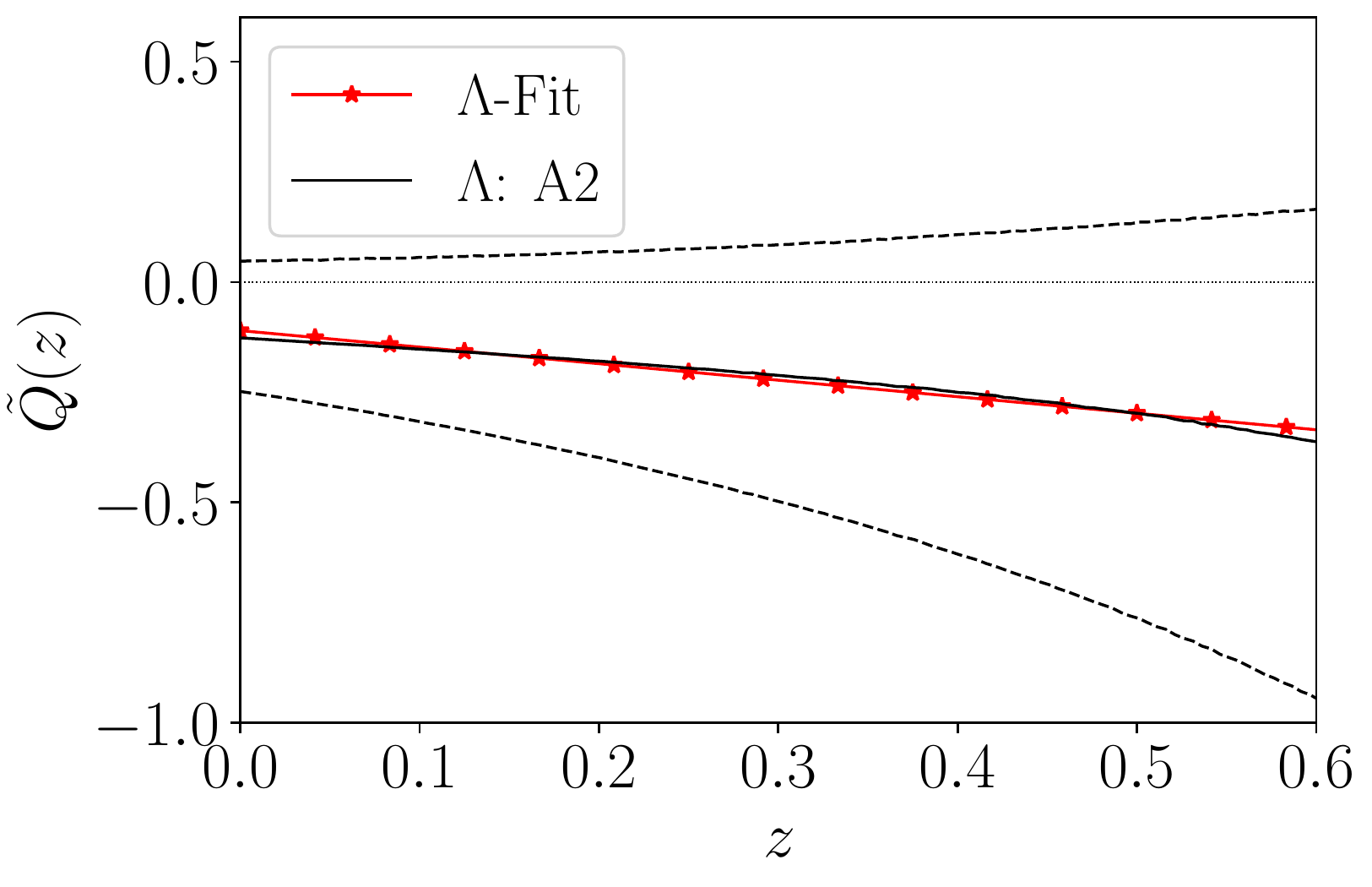}
		\includegraphics[angle=0, width=0.25\textwidth]{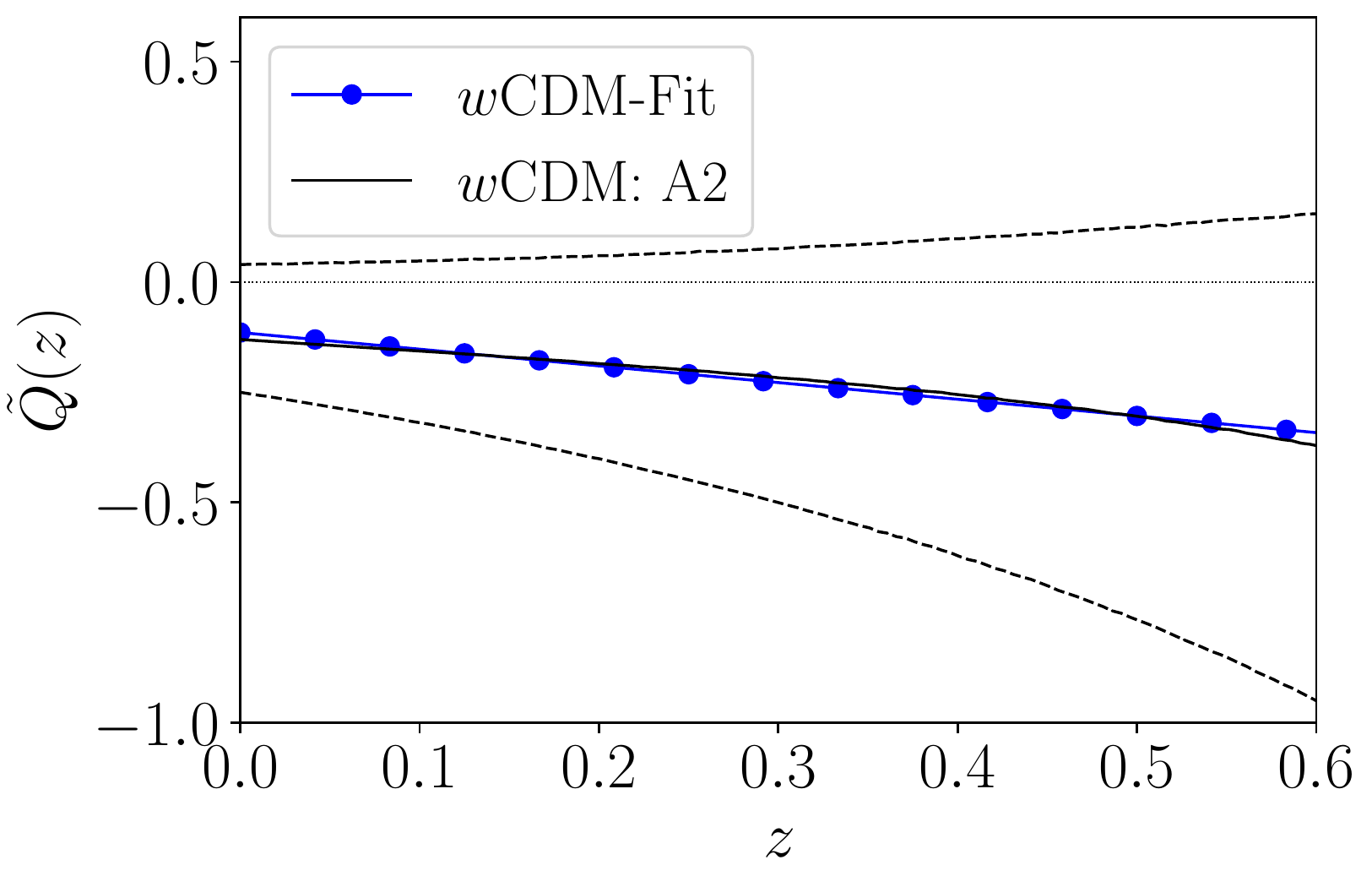}
		\includegraphics[angle=0, width=0.25\textwidth]{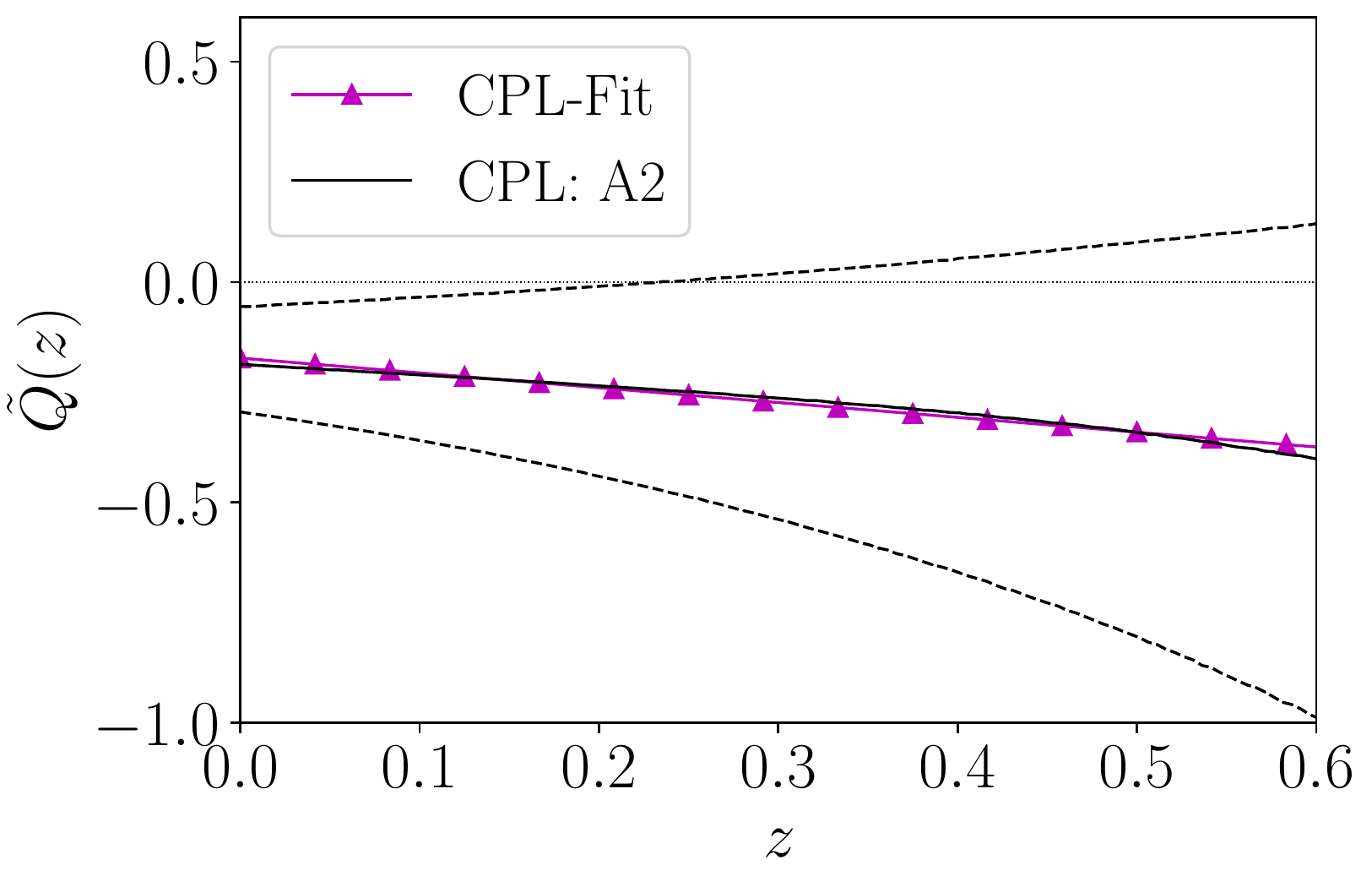}\\
		\includegraphics[angle=0, width=0.25\textwidth]{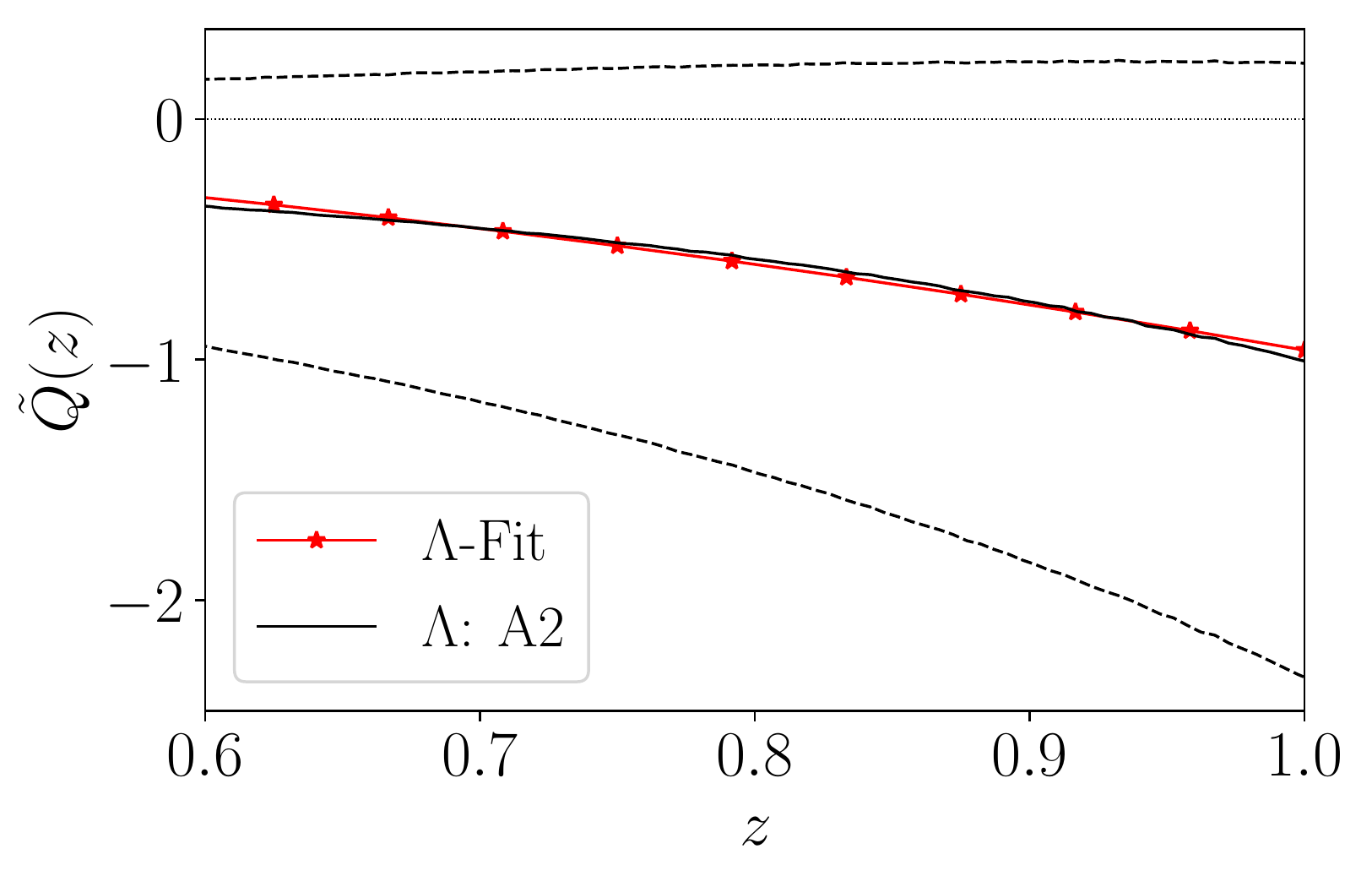}
		\includegraphics[angle=0, width=0.25\textwidth]{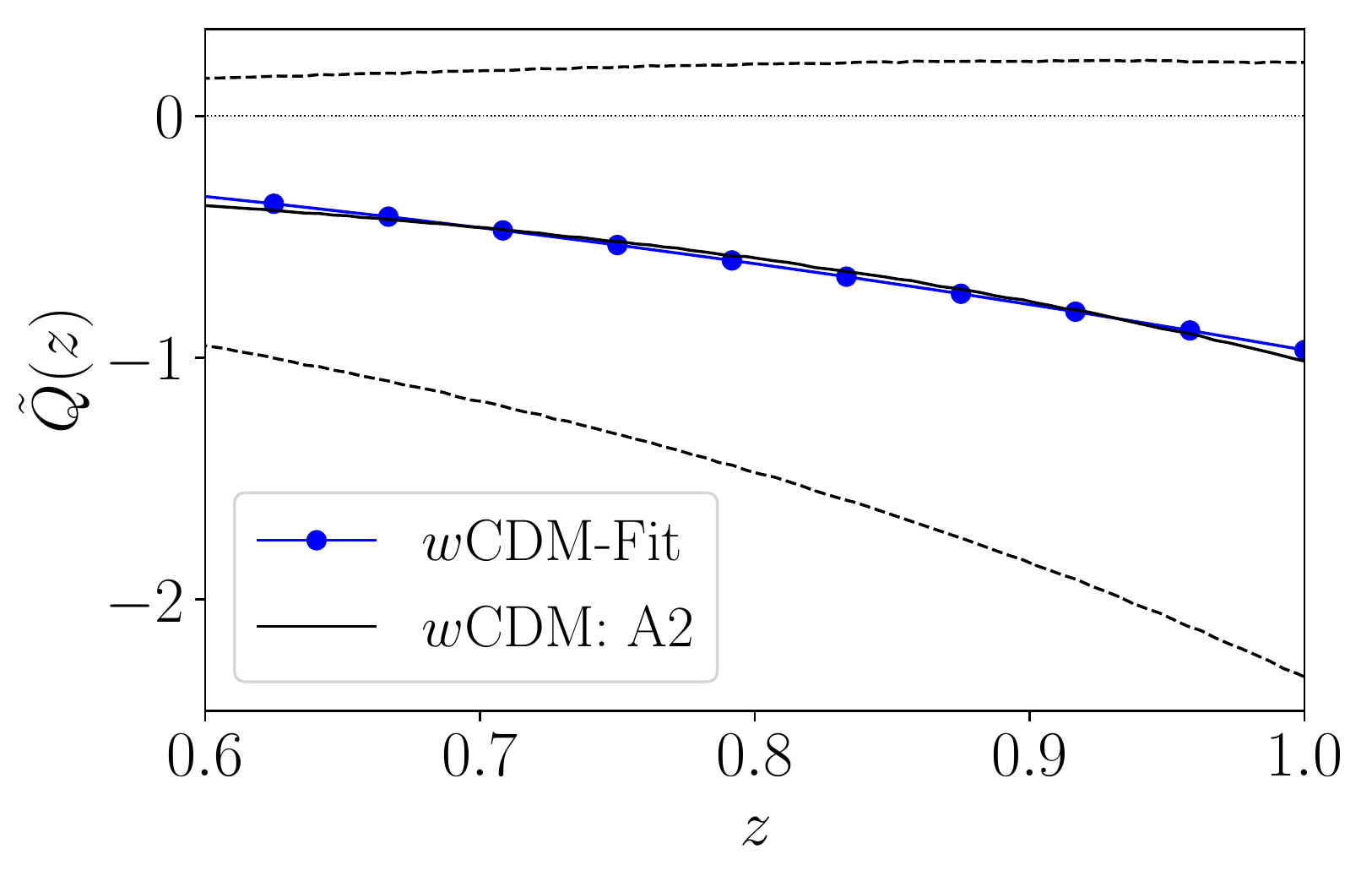}
		\includegraphics[angle=0, width=0.25\textwidth]{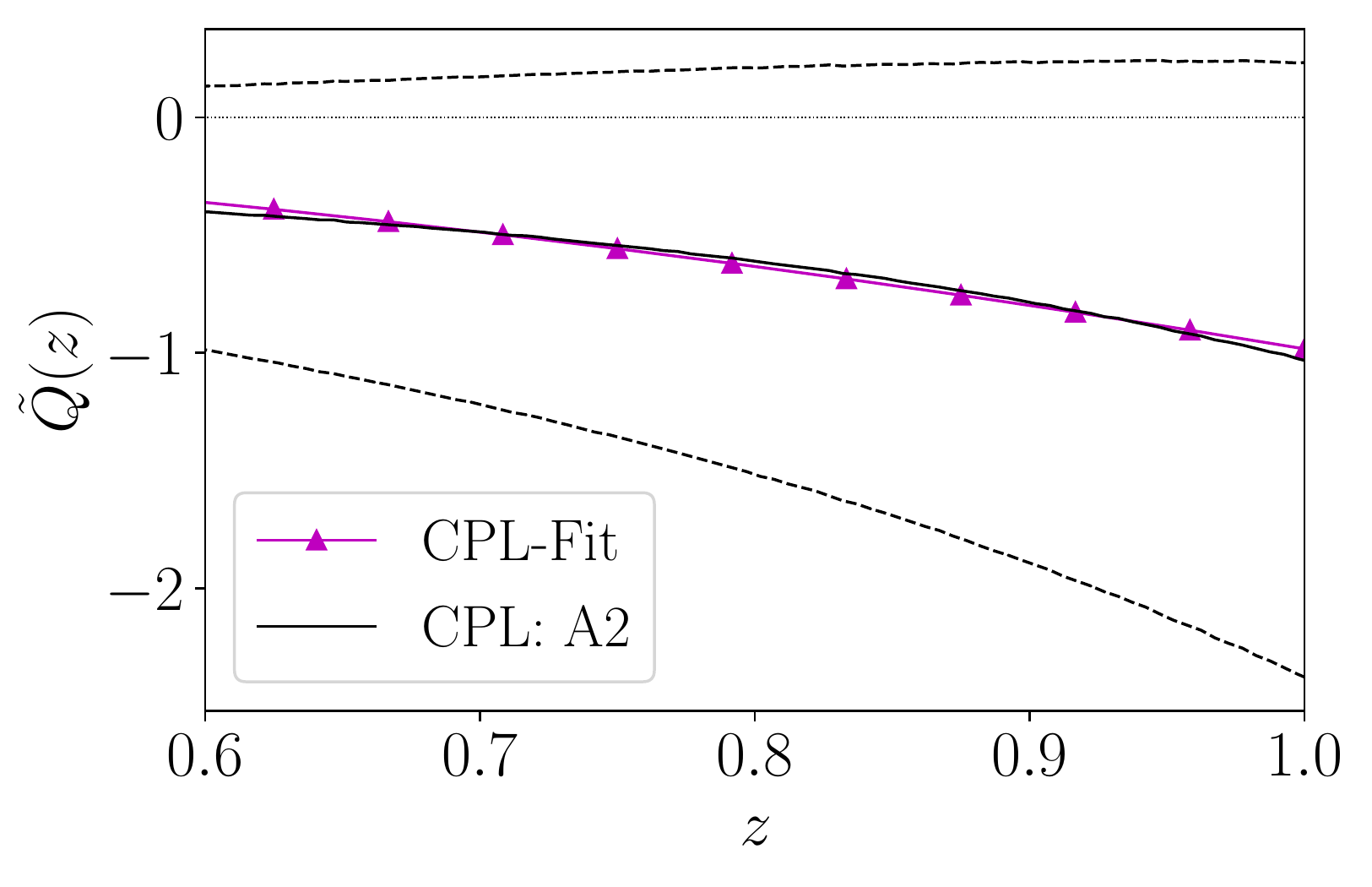}
	\end{center}
	\caption{{\small Plots showing a comparison between the reconstructed interaction $\tilde{Q}(z)$ and the estimated $\tilde{Q}_{\mbox{\tiny fit}}(z)$ using 
			the combined dataset A2, for EoS given by $w=-1$(left), $w$CDM model (middle) and the CPL parametrization (right). The black solid line is 
			the reconstructed function. The line with marker represents the best fit result from $\chi^2$-minimization. The 1$\sigma$ C.L.s are shown in dashed 
			lines.}}
	\label{Qfit_plot2}
\end{figure*}

\begin{figure*}
	\begin{center}
		\includegraphics[angle=0, width=0.25\textwidth]{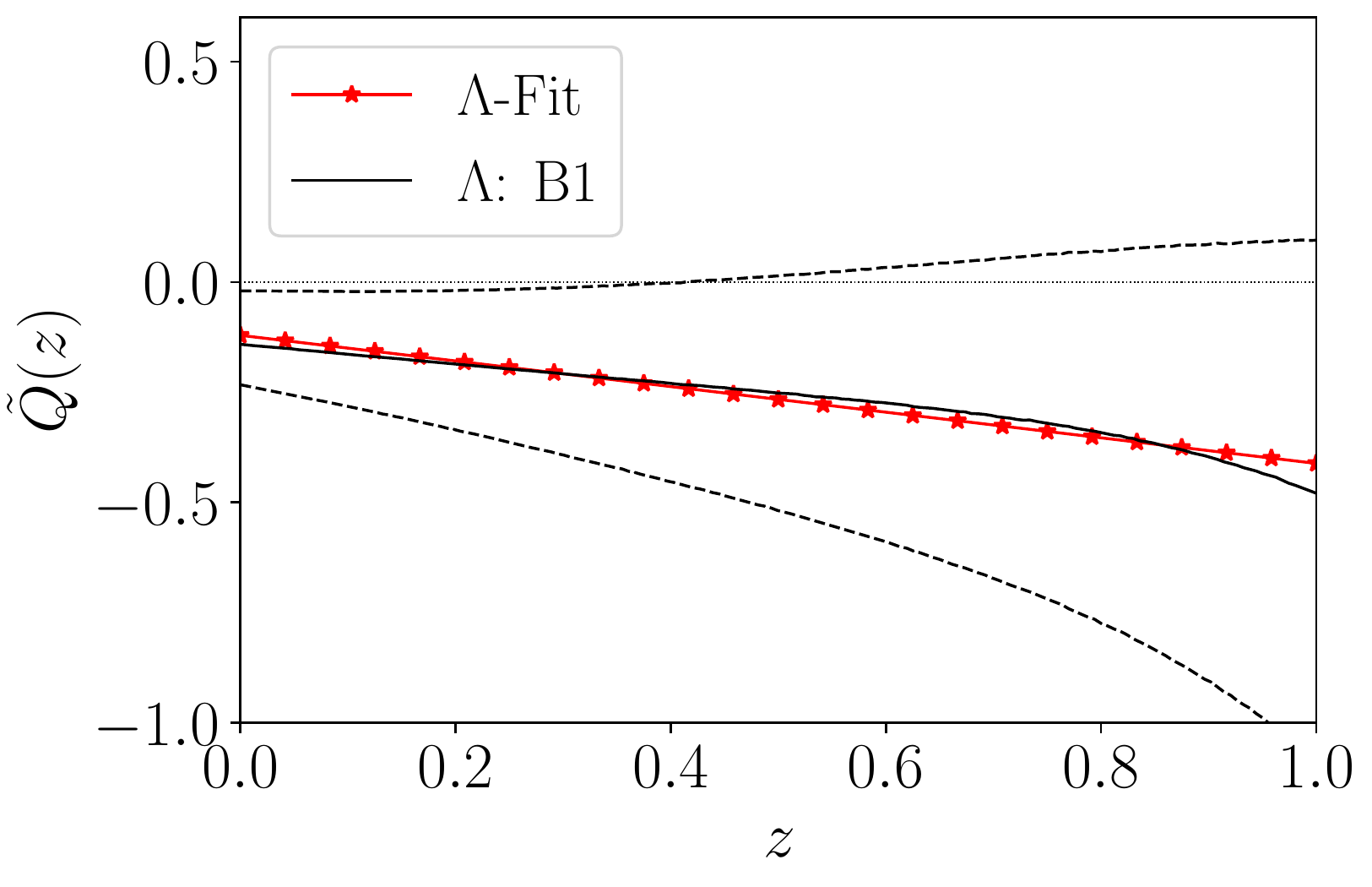}
		\includegraphics[angle=0, width=0.25\textwidth]{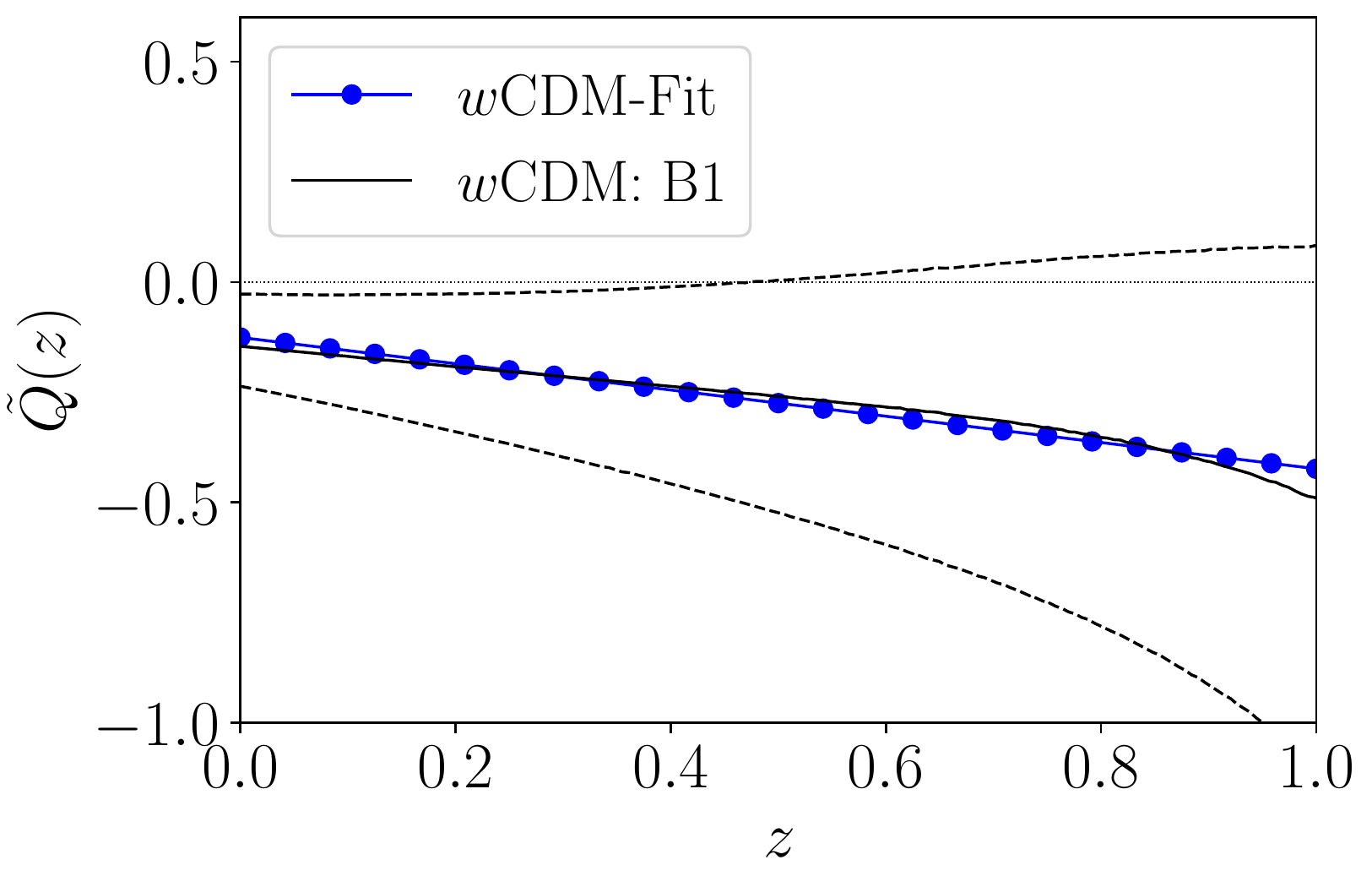}
		\includegraphics[angle=0, width=0.25\textwidth]{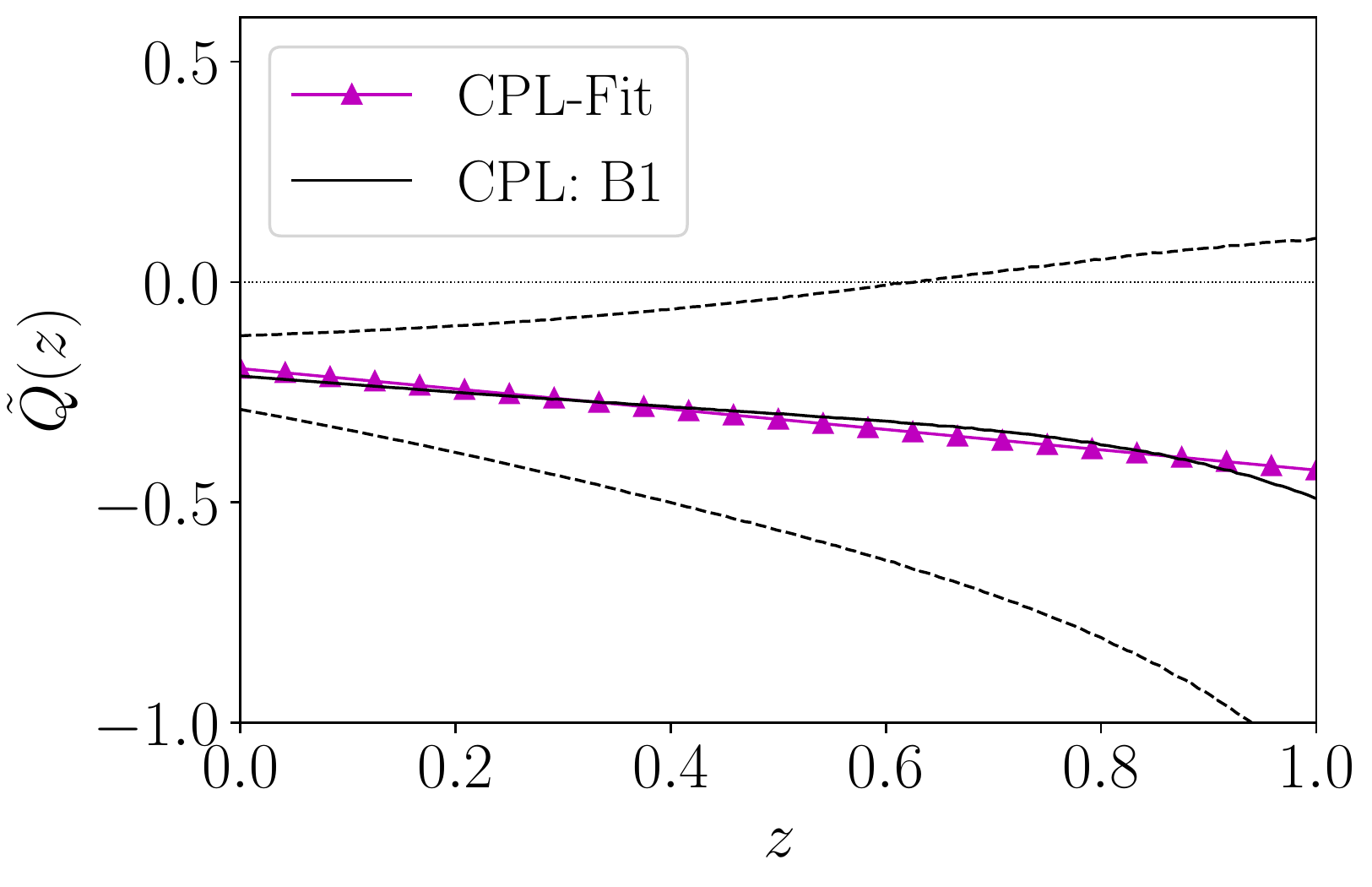}
	\end{center}
	\caption{{\small Plots showing a comparison between the reconstructed interaction $\tilde{Q}(z)$ and the estimated $\tilde{Q}_{\mbox{\tiny fit}}(z)$ using 
			the combined dataset B1, for EoS given by $w=-1$(left), $w$CDM model (middle) and the CPL parametrization (right). The black solid line is 
			the reconstructed function. The line with marker represents the best fit result from $\chi^2$-minimization. The 1$\sigma$ C.L.s are shown in dashed 
			lines.}}
	\label{Qfit_plot3}
\end{figure*}

\begin{figure*}
	\begin{center}
		\includegraphics[angle=0, width=0.25\textwidth]{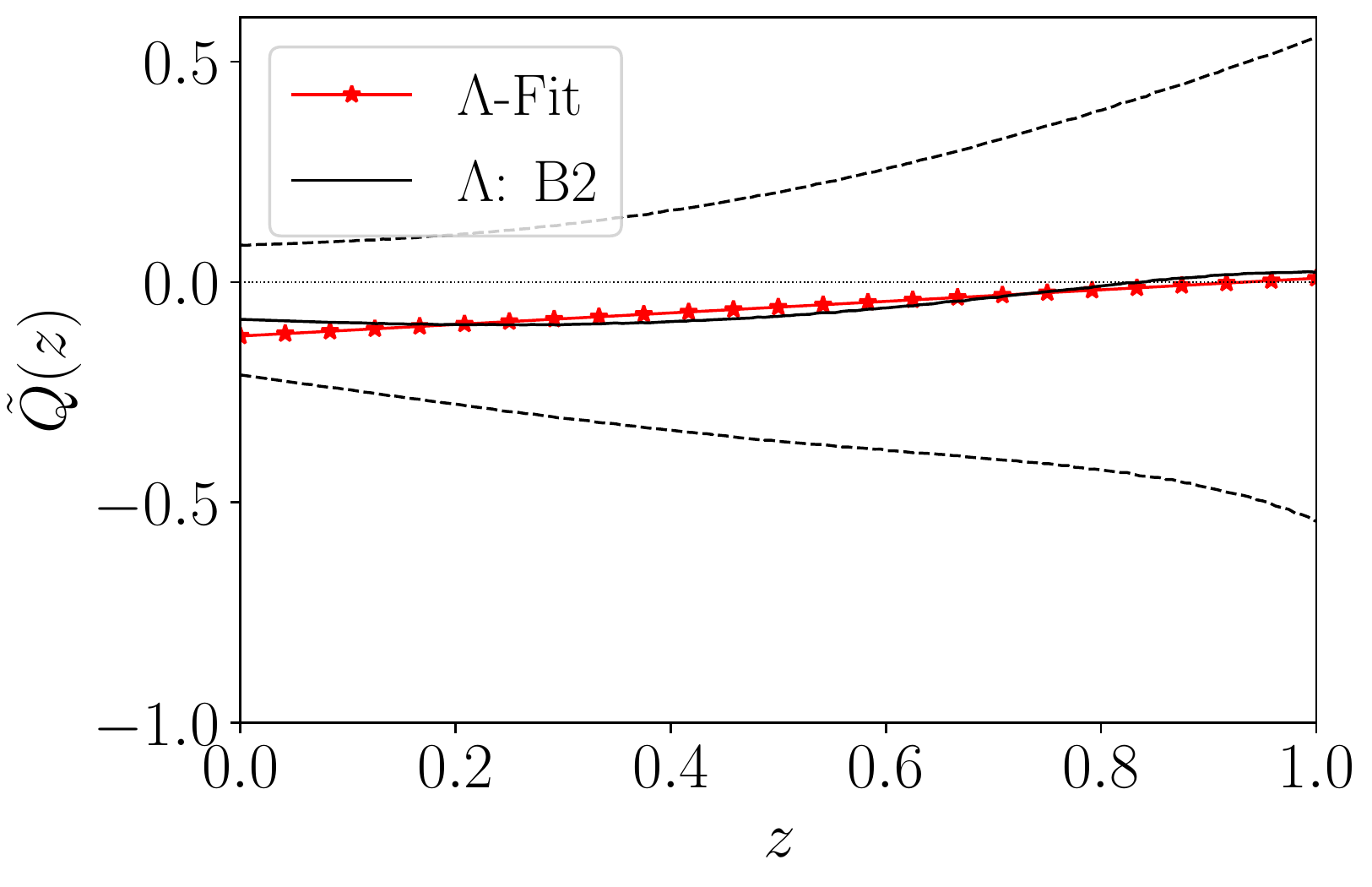}
		\includegraphics[angle=0, width=0.25\textwidth]{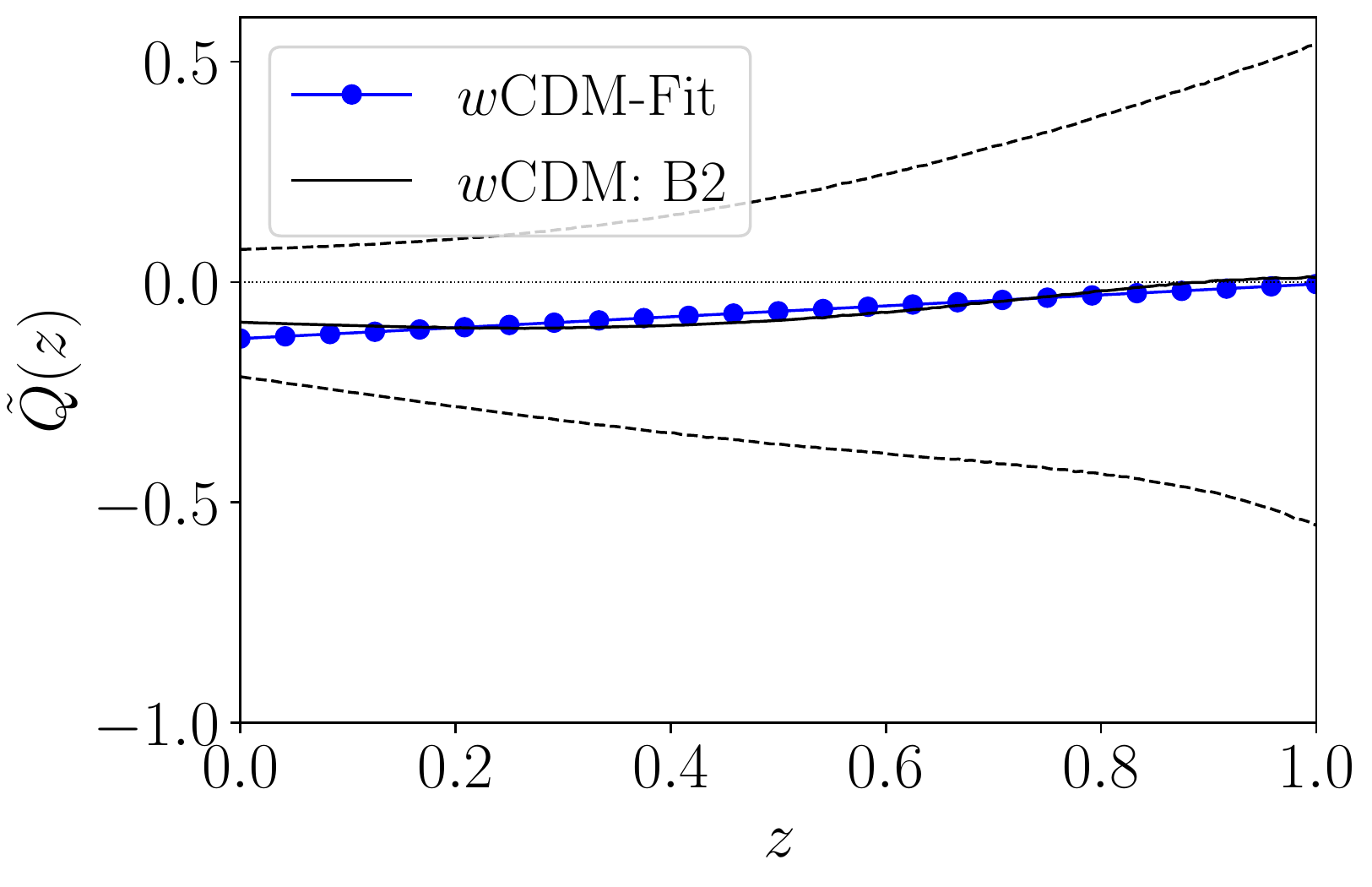}
		\includegraphics[angle=0, width=0.25\textwidth]{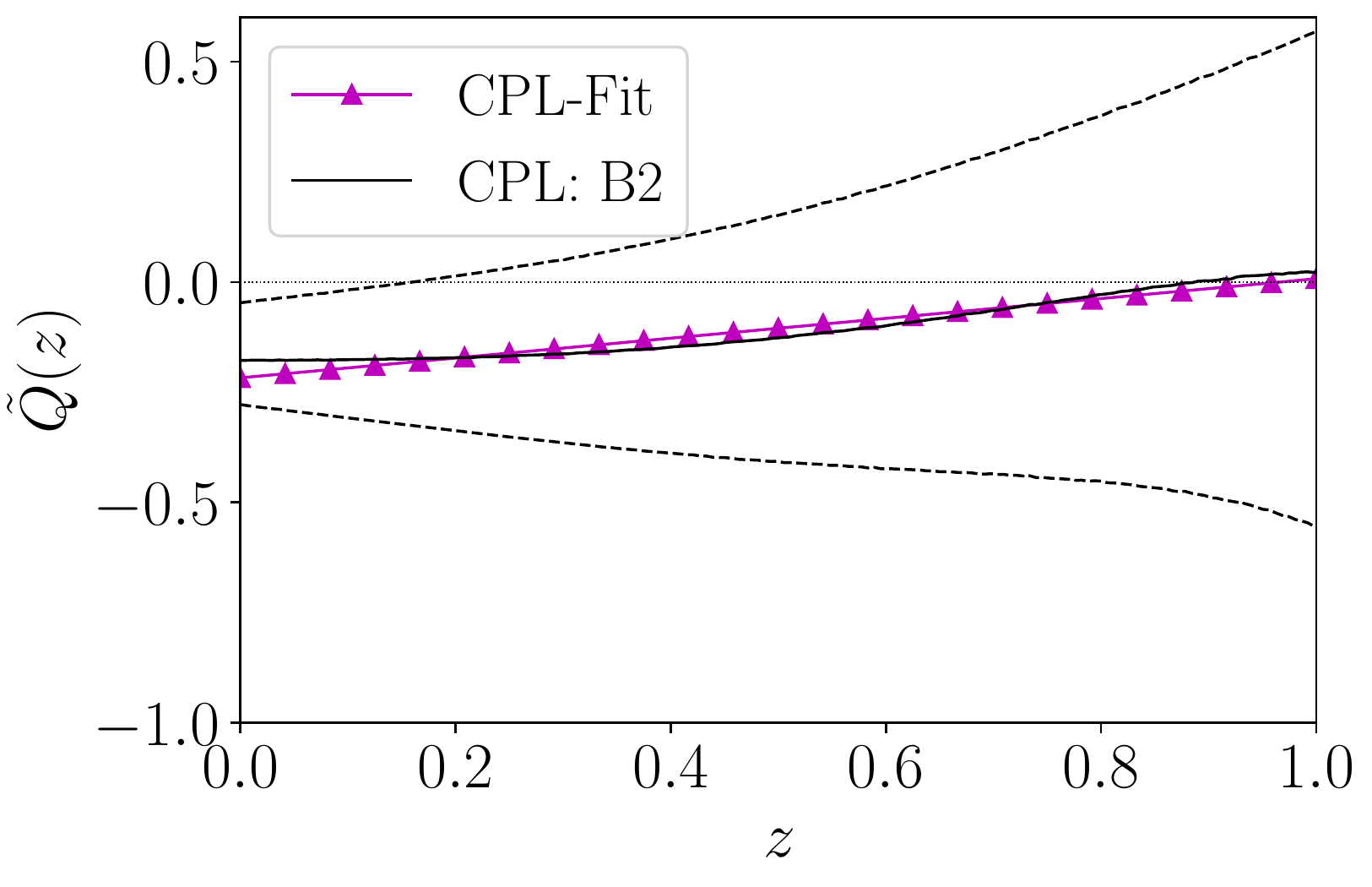}
	\end{center}
	\caption{{\small Plots showing a comparison between the reconstructed interaction $\tilde{Q}(z)$ and the estimated $\tilde{Q}_{\mbox{\tiny fit}}(z)$ using 
			the combined dataset B2, for EoS given by $w=-1$(left), $w$CDM model (middle) and the CPL parametrization (right). The black solid line is 
			the reconstructed function. The line with markers represents the best fit result from $\chi^2$-minimization. The 1$\sigma$ C.L.s are shown in dashed 
			lines.}}
	\label{Qfit_plot4}
\end{figure*}

\begin{table*}[t!] 
	\caption{{\small Table showing the coefficient $\tilde{Q}_i$'s for best fit $\tilde{Q}_{\mbox{\tiny fit}} = \tilde{Q}_0 + \tilde{Q}_1 z $ in the redshift 
			range $0<z<0.6$ and $\tilde{Q}_{\mbox{\tiny fit}} = \tilde{Q}_0 + \tilde{Q}_2 z^2 $ in the redshift range $0.6<z<1$ for datasets A1 and A2.}}
	\begin{center}
		\resizebox{0.999\textwidth}{!}{\renewcommand{\arraystretch}{1.25} \setlength{\tabcolsep}{12pt} \centering  
			\begin{tabular}{|c|c|c|c|c|c|c|} 
				\hline
				EoS & \textbf{Datasets} & $\tilde{Q}_0$ & $\tilde{Q}_1$ & \textbf{Datasets} & $\tilde{Q}_0$ & $\tilde{Q}_2$ \\
				\hline
				$w=-1$ & A1 $(0<z<0.6)$ &  $-0.060^{+0.180}_{-0.180}$ & $-0.322^{+0.516}_{-0.514}$ & A1 $(0.6<z<1)$ &  $0.443^{+0.329}_{-0.383}$ &  $-1.865^{+0.558}_{-0.481}$  \\ 
				\hline				
				$w$CDM & A1 $(0<z<0.6)$ & $-0.063^{+0.180}_{-0.180}$ & $-0.324^{+0.516}_{-0.516}$ & A1 $(0.6<z<1)$ & $0.438^{+0.332}_{-0.383}$ &  $-1.866^{+0.558}_{-0.484}$   \\ 
				\hline
				CPL & A1 $(0<z<0.6)$ & $-0.114^{+0.180}_{-0.180}$ & $-0.286^{+0.515}_{-0.518}$ & A1 $(0.6<z<1)$ & $0.434^{+0.336}_{-0.385}$  &  $-1.883^{+0.561}_{-0.491}$  \\ 
				\hline
				$w=-1$ & A2 $(0<z<0.6)$ & $-0.110^{+0.180}_{-0.180}$ & $-0.375^{+0.517}_{-0.516}$ & A2 $(0.6<z<1)$ &  $0.031^{+0.400}_{-0.401}$ &  $-0.992^{+0.586}_{-0.583}$  \\ 
				\hline				
				$w$CDM & A2 $(0<z<0.6)$ & $-0.115^{+0.179}_{-0.180}$ & $-0.379^{+0.517}_{-0.516}$ & A2 $(0.6<z<1)$ & $0.025^{+0.401}_{-0.401}$ &  $-0.993^{+0.586}_{-0.583}$ \\ 
				\hline
				CPL & A2 $(0<z<0.6)$ & $-0.173^{+0.180}_{-0.179}$ & $-0.336^{+0.517}_{-0.517}$ & A2 $(0.6<z<1)$ & $-0.011^{+0.401}_{-0.400}$  &  $-0.973^{+0.582}_{-0.584}$  \\ 
				\hline
		\end{tabular}}
	\end{center}
	\label{qfit_tabA}
\end{table*}

\begin{table*}[t!] 
	\caption{{\small Table showing the coefficient $\tilde{Q}_i$'s for best fit $\tilde{Q}_{\mbox{\tiny fit}} = \tilde{Q}_0 + \tilde{Q}_1 z $ in the redshift 
			range $0<z<1$ for datasets B1 and B2.}}
	\begin{center}
		\resizebox{0.999\textwidth}{!}{\renewcommand{\arraystretch}{1.25} \setlength{\tabcolsep}{12pt} \centering  
			\begin{tabular}{|c|c|c|c|c|c|c|} 
				\hline
				EoS & \textbf{Datasets} & $\tilde{Q}_0$ & $\tilde{Q}_1$ & \textbf{Datasets} & $\tilde{Q}_0$ & $\tilde{Q}_1$ \\
				\hline
				$w=-1$ & B1 $(0<z<1)$ &  $-0.121^{+0.153}_{-0.153}$ &  $-0.291^{+0.262}_{-0.263}$ & B2 $(0<z<1)$ &  $-0.122^{+0.153}_{-0.153}$ &  $0.131^{+0.263}_{-0.263}$  \\ 
				\hline				
				$w$CDM & B1 $(0<z<1)$ & $-0.126^{+0.152}_{-0.153}$ &  $-0.298^{+0.264}_{-0.262}$ & B2 $(0<z<1)$ & $-0.128^{+0.153}_{-0.153}$ &  $0.124^{+0.264}_{-0.263}$ \\ 
				\hline
				CPL & B1 $(0<z<1)$ & $-0.196^{+0.153}_{-0.153}$ &  $-0.231^{+0.263}_{-0.263}$ & B2 $(0<z<1)$ & $-0.217^{+0.153}_{-0.153}$  &  $0.225^{+0.263}_{-0.263}$ \\ 
				\hline
			\end{tabular}
		}
	\end{center}
	\label{qfit_tabB}
\end{table*}

In the present work, we make an attempt to get an essence of how the reconstructed interaction function evolves 
with redshift. The motivation is to find the nature of deviation from the zero interaction scenario. We reconstruct the dimensionless cosmic interaction $\tilde{Q}(z)$ using the 
GP method for three choices of the dark energy equation of state parameter $w$. Also, two choices for the covariance function have been considered. The reconstructed interaction 
function $\tilde{Q}$ for various combinations of datasets for each of the three dark energy models for the two choices of covariance function are shown in Fig. \ref{Q_sqexp} 
and \ref{Q_mat92}. The shaded regions correspond to the $68\%$, $95\%$ and $99.7\%$ confidence levels (CL) respectively from darker to lighter shades. The black solid line 
shows the curve with best-fit values of $\tilde{Q}$. Table \ref{Q_lcdm_res}, \ref{Q_wcdm_res} and \ref{Q_cpl_res} show the best fit results for 
$\tilde{Q}(z=0)$ along with the $1\sigma$, 2$\sigma$ and 3$\sigma$ uncertainties for all the combinations. In Fig. \ref{Qscaled_sqexp} and \ref{Qscaled_mat92}, we zoom in the 
plots for $\tilde{Q}$ in the range $0<z<0.5$, to look at its behaviour at very low redshift more closely. It may be noted that all these figures, \ref{Q_sqexp} to 
\ref{Qscaled_mat92}, are essentially the plots of the equation \eqref{Q_h}.  \\

From equation \eqref{conservation} one can understand that a negative $Q$ indicates the energy flow from dark energy to dark matter sector, and a positive $Q$ indicates the 
reverse. The plots show that the interaction function $\tilde{Q}$ remains close to $0$ indicating no appreciable interaction for low redshift ranges. The best fit curve shows 
a small deviation towards negative values, but the zero interaction scenario is always included in 2$\sigma$ for most of the combinations. So the energy gets transferred from 
the dark energy to the dark matter sector if it happens at all. This direction of flow of energy is consistent with the thermodynamic 
requirement as discussed by Pav\'{o}n and Wang \cite{diego}. Interestingly, the case of $\tilde{Q} < 0$ guarantees that the ratio $\frac{\rho_{m}}{\rho_D}$ asymptotically 
tends to a constant \cite{pavon}, and thus alleviaties the coincidence problem.

\section{Fitting function for $\tilde{Q}$}

In this section, an approximate fitting formula for the reconstructed interaction is derived. This is done in the low redshift range $0 < z < 1$ using the combined datasets
A1, A2, B1 and B2. The goal is to find a simple analytic form of $\tilde{Q}$ prior to the transition. As both the covariance functions yield similar results, we pick up the 
cases for one of them, namely the Mat\'{e}rn 9/2 covariance as the example. We consider a polynomial for $\tilde{Q}(z)$ as a function of redshift $z$ as,
\begin{equation} \label{Qfit}
\tilde{Q}_{\mbox{\tiny fit}} (z) = \sum_{i=0}^{n} \tilde{Q}_i z^i.
\end{equation}

We estimate the coefficients $\tilde{Q}_i$'s of the above equation by the $\chi^2$ minimization, where we define the $\chi^2$ function as,
\begin{equation}
\chi^2 = \sum_{s}\dfrac{\left[\tilde{Q}(z_s)- \tilde{Q}_{\mbox{\tiny fit}}(z_s)\right]^2}{\sigma^2(z_s)}.
\end{equation}

We perform the fitting using a trial and error estimation for different orders of $i$ in equation \eqref{Qfit}. The value of the reduced $\chi^2$, defined as 
$\chi^2_\nu = \frac{\chi^2}{\nu}$, 
where $\nu$ signifies the degrees of freedom, is estimated. This procedure entails to go from order to order in the polynomial and getting the best-fitting $\chi^2$, 
and truncating once $\chi_\nu^2$ falls below one to prevent over-fitting. Again, the estimated $\tilde{Q}_i$'s along with their $1\sigma$ uncertainties are given. 
A comparison between the reconstructed $\tilde{Q}(z)$ and estimated $\tilde{Q}_{\mbox{\tiny fit}}$, for various combinations of datasets are shown in Figures \ref{Qfit_plot1}, 
\ref{Qfit_plot2}, \ref{Qfit_plot3} and \ref{Qfit_plot4}. The best fit values of the coefficients are shown in Tables \ref{qfit_tabA} and \ref{qfit_tabB}. \\

\begin{figure*} [t!]
	\begin{center}
		\includegraphics[angle=0, width=\textwidth]{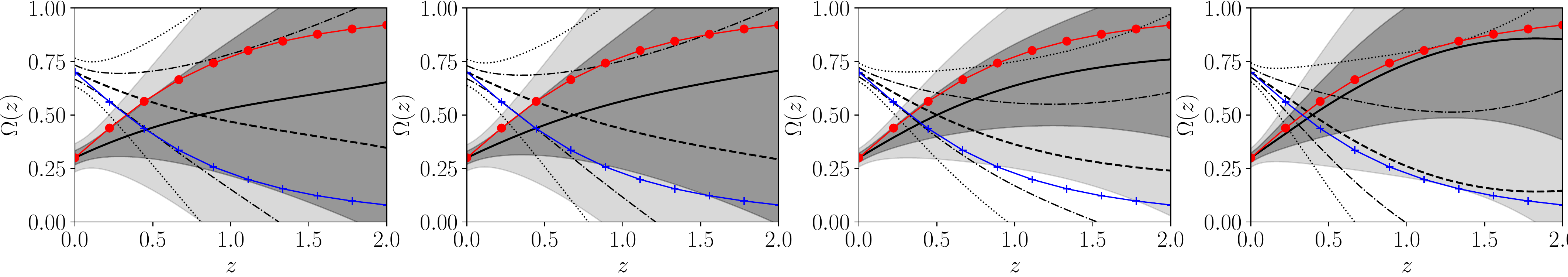}\\
		\includegraphics[angle=0, width=\textwidth]{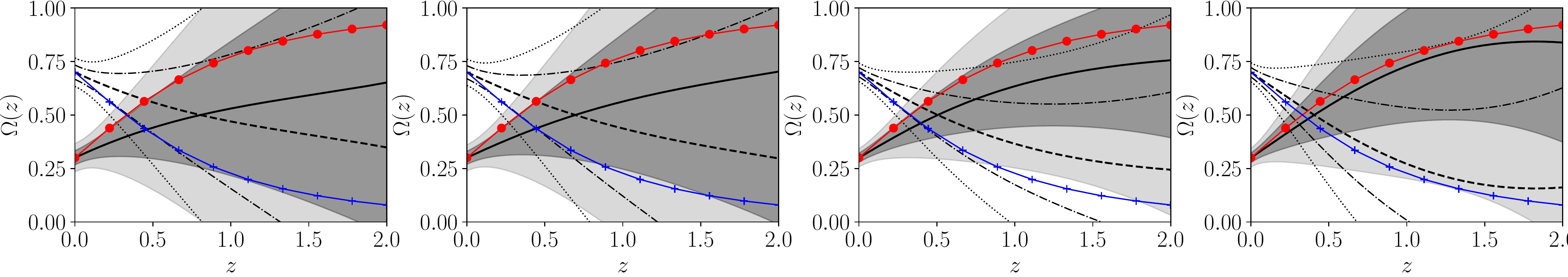}\\
		\includegraphics[angle=0, width=\textwidth]{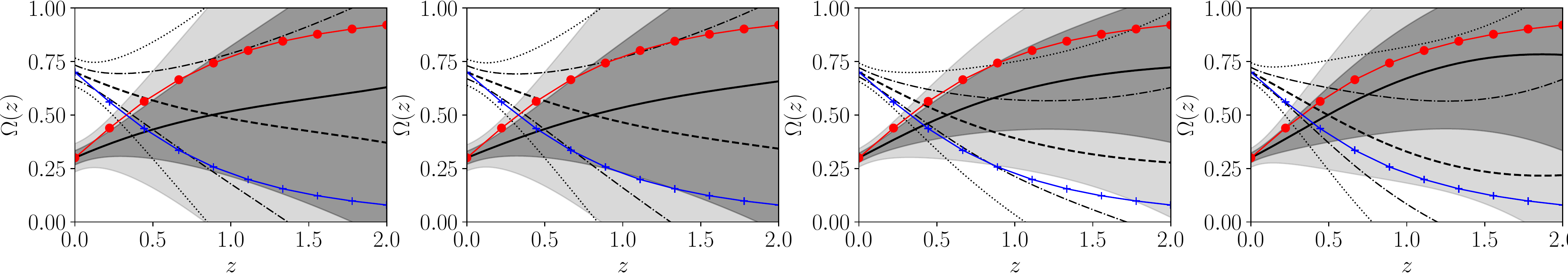}
	\end{center}
	\caption{{\small Plots for the dark energy density $\Omega_D$ and the matter density $\Omega_{m}$ from the dataset samples of Set A1 (column 1), Set A2 (column 2), 
			Set B1 (column 3) and Set B2 (column 4) using a Squared Exponential Covariance function considering the decaying Dark Energy EoS given by $w = -1$ (top row), the 
			$w$CDM model with DE EoS given by $w = -1.006 \pm 0.045$\cite{wcdm} (middle row), and the CPL parametrization of Dark Energy with EoS given by 
			$w(z) = w_0 + w_a (\frac{z}{1+z})$, $w_0 = -1.046^{+0.179}_{-0.170}$ and $w_a = 0.14^{+0.60}_{-0.76}$\cite{cpl} (bottom row). The black solid curve corresponds to 
			$\Omega_m$ while the black dashed line represents $\Omega_{D}$. The 1$\sigma$ and 2$\sigma$ uncertainties in $\Omega_{m}$ is shown by the dark and light shaded regions, 
			and those of $\Omega_{D}$ is given by the region bounded with dashed-dotted and dotted lines respectively. The line drawn with circles represents $\Omega_{m}$ and the 
			line with cross markers is that of $\Omega_{D}$, for the $\Lambda$CDM model.}}
	\label{omega_sqexp}
\end{figure*}

\begin{figure*} 
	\begin{center}
		\includegraphics[angle=0, width=\textwidth]{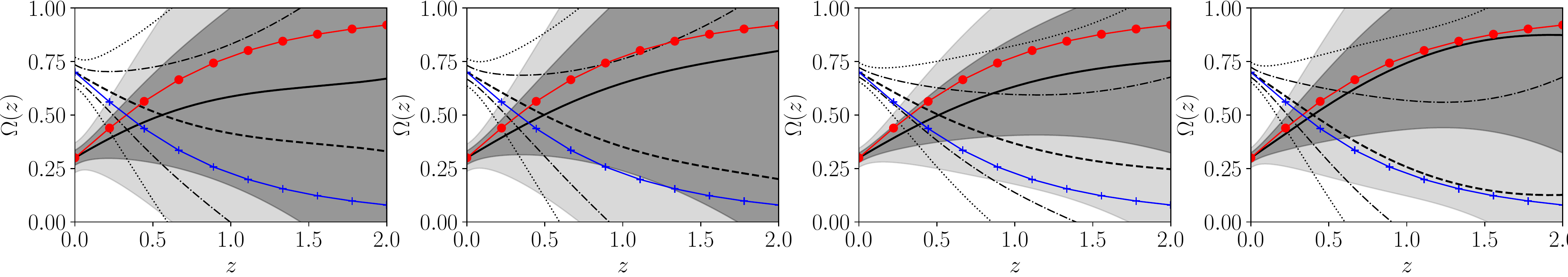}\\
		\includegraphics[angle=0, width=\textwidth]{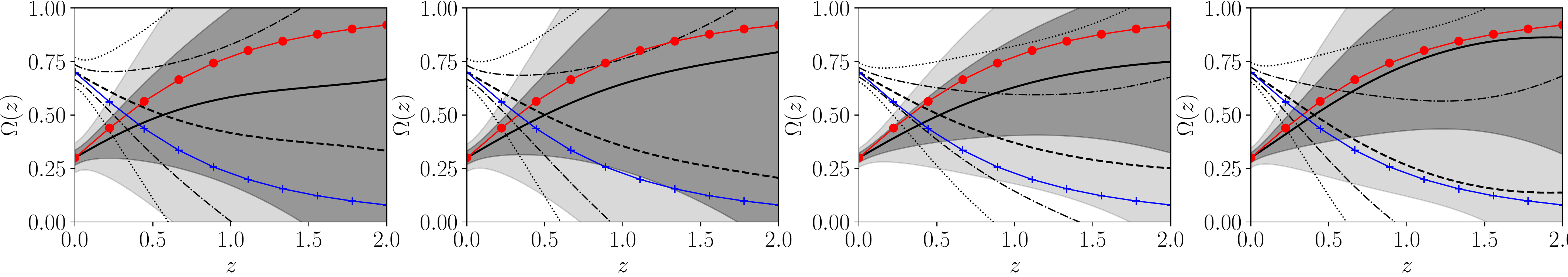}\\
		\includegraphics[angle=0, width=\textwidth]{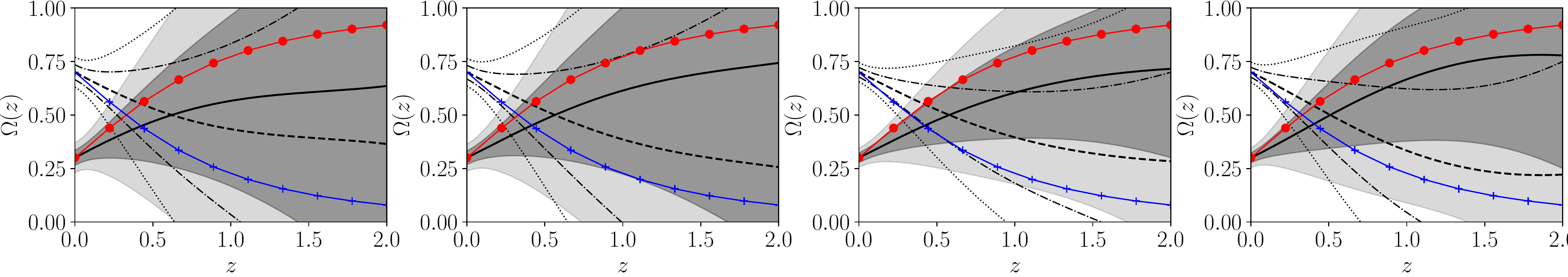}
	\end{center}
	\caption{{\small Plots for the dark energy density $\Omega_D$ and the matter density $\Omega_{m}$ from the dataset samples of Set A1 (column 1), Set A2 (column 2), 
			Set B1 (column 3) and Set B2 (column 4) using a Mat\'{e}rn 9/2 Covariance function considering the decaying Dark Energy EoS given by $w = -1$ (top row), the $w$CDM 
			model with DE EoS given by $w = -1.006 \pm 0.045$\cite{wcdm} (middle row), and the CPL parametrization of Dark Energy with EoS given by $w(z) = w_0 + w_a (\frac{z}{1+z})$, 
			$w_0 = -1.046^{+0.179}_{-0.170}$ and $w_a = 0.14^{+0.60}_{-0.76}$\cite{cpl} (bottom row). The black solid curve corresponds to $\Omega_m$ while the black dashed line 
			represents $\Omega_{D}$. The 1$\sigma$ and 2$\sigma$ uncertainties in $\Omega_{m}$ is shown by the dark and light shaded regions, and that of $\Omega_{D}$ is given by the 
			region bounded with dashed-dotted and dotted lines respectively. The line drawn with circles represents $\Omega_{m}$ and the line with cross markers is that of 
			$\Omega_{D}$, for the $\Lambda$CDM model.}}
	\label{omega_mat92}
\end{figure*}

For A1 dataset, in the redshift range $0<z<0.6$,
\begin{eqnarray}
\tilde{Q}_{\mbox{\tiny fit}}(z) &=& -0.060 -0.322~z   \mbox{ for $w=-1$} .\\
 &=& -0.063 -0.324~z   \mbox{ for $w$CDM} .\\
 &=& -0.114 -0.286~z  \mbox{ for CPL} .
\end{eqnarray} 

For A1 dataset, in the redshift range $0.6<z<1$,
\begin{eqnarray}
\tilde{Q}_{\mbox{\tiny fit}}(z) &=& 0.443 -1.865~z^2  \mbox{ for $w=-1$} .\\
&=& 0.438 -1.866~z^2  \mbox{ for $w$CDM} .\\
&=& 0.424 -1.882~z^2  \mbox{ for CPL} .
\end{eqnarray} 

For A2 dataset, in the redshift range $0<z<0.6$,
\begin{eqnarray}
\tilde{Q}_{\mbox{\tiny fit}}(z) &=& -0.110 -0.375~z  \mbox{ for $w=-1$} .\\
&=& -0.115 -0.379~z  \mbox{ for $w$CDM} .\\
&=& -0.173 -0.336~z \mbox{ for CPL} .
\end{eqnarray} 

For A2 dataset, in the redshift range $0.6<z<1$,
\begin{eqnarray}
\tilde{Q}_{\mbox{\tiny fit}}(z) &=& 0.031 -0.992~z^2  \mbox{ for $w=-1$} .\\
&=& 0.025 -0.993~z^2  \mbox{ for $w$CDM} .\\
&=& -0.011 -0.973~z^2  \mbox{ for CPL} .
\end{eqnarray} 

For B1 dataset, in the redshift range $0<z<1$,
\begin{eqnarray}
\tilde{Q}_{\mbox{\tiny fit}}(z) &=& -0.121 -0.291~z \mbox{ for $w=-1$} .\\
&=& -0.126 -0.298~z \mbox{ for $w$CDM} .\\
&=& -0.196 -0.231~z \mbox{ for CPL} .
\end{eqnarray} 

For B2 dataset, in the redshift range $0<z<1$,
\begin{eqnarray}
\tilde{Q}_{\mbox{\tiny fit}}(z) &=& -0.122 +0.131~z \mbox{ for $w=-1$} .\\
&=& -0.128 +0.124~z \mbox{ for $w$CDM} .\\
&=& -0.217 -0.225~z \mbox{ for CPL} .
\end{eqnarray} 

If we proceed on fitting with any higher order polynomial, the fitted function will not be contained within the $1\sigma$ error margin of 
$\tilde{Q}(z)$ reconstructed by GP.

\section{Evolution of the cosmological density parameters}

With the nature of the interaction function reconstructed, one can obtain the evolution of energy density parameters as well. 
The model is a spatially flat, homogenous and isotropic universe 
where the total energy density is composed of only a pressureless matter and the dark energy. We define the density parameters $\Omega_i$'s as,
\begin{eqnarray} 
\Omega_{m} &=& \frac{\tilde{\rho}_m}{E^2}, \label{omega_m}\\
\Omega_{D} &=& \frac{\tilde{\rho}_D}{E^2}, \label{omega_d}
\end{eqnarray} 
such that $\Omega_{m} + \Omega_{D} = 1$.\\

We make use of the equation \eqref{cons_reduced_m} and rewrite it as, 
\begin{eqnarray} 	
\frac{d\tilde{\rho}_m}{dz} - \frac{3 \tilde{\rho}_m}{1+z} &=& \frac{\tilde{Q}}{E (1+z)}. \label{rho_m}
\end{eqnarray}
One can see that \eqref{rho_m} is a linear first order non-homogeneous differential equation of the form, 
\begin{equation}\label{lin1DE}
\frac{dy}{dz} + A(z)y = B(z)
\end{equation} 
with $A(z) = -\frac{3}{1+z}$, and $B(z) = \frac{\tilde{Q}}{E (1+z)}$. The integrating factor for \eqref{lin1DE} is given by $e^{\int A(z)dz}$, 
and the general solution is 
\begin{equation}\label{lin1DEsol}
	y = e^{-\int A dz} \int\left(B e^{\int A dz} \right) dz + C,
\end{equation} 
where $C$ is the constant of integration. \\

Similarly, we solve for equation \eqref{rho_m} and arrive at an expression for the reduced matter density $\tilde{\rho}_m$ as,
\begin{eqnarray} \label{rho_msol}
	\tilde{\rho}_m = 	\tilde{\rho}_{m0}(1+z)^{3} + {(1+z)^{3}}\int_{0}^{z}\frac{\tilde{Q}}{E } (1+z)^{-4} dz.
\end{eqnarray} 
If $\tilde{Q} = 0$, equation \eqref{rho_msol} reduces to the relation $\tilde{\rho}_{m} = \tilde{\rho}_{m0}(1+z)^3$ (where $\tilde{\rho}_{m0} = \tilde{\rho}_{m} (z=0)$) 
which is the standard evolution scenario for a pressureless matter that evolves independently. One can  
now write the density parameters with the help of the equations \eqref{omega_m}-\eqref{omega_d} as, 

\begin{eqnarray} \label{omega_msol}
\Omega_{m} &=& 	\frac{\tilde{\rho}_{m0}(1+z)^{3}}{E^2} + \frac{(1+z)^{3}}{E^2}\int_{0}^{z}\frac{\tilde{Q}}{E (1+z)^{4}} ~ dz,~~~~~\\
\Omega_{D} &=& 1 - \Omega_{m}.
\end{eqnarray}

With the smooth functions of $E(z)$ and $\tilde{Q}(z)$ reconstructed from the combined datasets, we use the trapezoidal rule \cite{trapez} to calculate the integral
\begin{eqnarray} \label{f_gp}
f(z) &=& \int_{0}^{z}\frac{\tilde{Q}}{E } (1+z)^{-4} ~dz \nonumber \\
 &=& \int_{0}^{z} g(z)~ dz \nonumber \\
&\simeq& \frac{1}{2} \sum_{i=0}^{n} (z_{i+1} - z_i) \left[{g(z_{i+1})}+{g(z_{i})}\right],
\end{eqnarray} 
where $g(z) = \frac{\tilde{Q}}{E } (1+z)^{-4}$. The uncertainty in $f(z)$ is obtained by error propagation formula,
\begin{equation} \label{sigf_gp}
\sigma^2_f =  \frac{1}{4}  \sum_{i=0}^{n}\left(z_{i+1} - z_i\right)^2 \left[\sigma^2_{g_{i+1}}+\sigma^2_{g_{i}}\right],
\end{equation} 
where contribution from uncertainties in $\tilde{Q}$ and $E$ have been included.\\

We plot the density parameters $\Omega_{m}$ and $\Omega_{D}$ using the equations \eqref{omega_m} and \eqref{omega_d}. We choose the value of 
$\tilde{\rho}_{m0} = 0.3$, i.e., $\Omega_{m0} = \frac{0.3}{E(0)^2}$. The plots are shown in Fig. \ref{omega_sqexp} and 
Fig. \ref{omega_mat92} for the two choices of the covariance function. \\

For the three different choices of the interacting dark energy models, the evolution of the density parameters are found to be qualitatively similar and also not too
sensitive to the choice of the datasets. This feature hardly depends on the choice of the covariance function, only except the fact the use of Mat\'{e}rn $9/2$
covariance function brings the transition to dark energy dominance a bit closer to $z=0.5$. One intriguing common feature to note is that 
for the interacting models, $\Omega_D$ takes over as the dominant role over $\Omega_m$ later in the evolution (closer to $z=0$) compared to the corresponding $\Lambda$CDM 
model. \\

One can note that the transition from a matter dominated phase to a dark energy dominated phase occurs within the redshift range $0.5<z<1$. \\

\section{Thermodynamics of the Interaction}

For the thermodynamic properties of the model, we consider the universe as a system that is bounded by some cosmological 
horizon, and the matter content of the universe is enclosed within a volume defined by a radius not bigger than the horizon. It deserves mention 
that this idea primarily originated from the consideration of black hole thermodynamics, which are equally valid for a cosmological 
horizon\cite{gibbons, jacob, padma}. However, in an evolving scenario such as in cosmology, an apparent horizon is more relevant than
an event horizon. An apparent horizon is given by the equation $g^{\mu\nu} R_{,\mu} R_{,\nu} = 0$. For a spatially flat FLRW universe, 
this equation tells us that the apparent horizon ($r_h$) is in fact the Hubble horizon, 
\begin{equation}
	r_h = \frac{1}{H}.
\end{equation}
This serves the purpose for recovering the first law of thermodynamics. For a comprehensive description, we refer to the work of Ferreira and Pav\'{o}n\cite{diego2}, and the 
monograph by Faraoni\cite{valerio}. \\

For the second law to be valid, the entropy $S$ should be non-decreasing w.r.t. the expansion of the universe. If $S_f$, $S_h$ stand for entropy of the fluid and 
that of the horizon containing the fluid respectively, then the total entropy of the system, i.e. $S = S_f + S_h$ , should satisfy the relation 
\begin{equation}\label{cond1}
 \frac{dS}{dx} \geq 0,
\end{equation} where $x = \ln a$, $a$ being the scale factor of the universe. For an approach to equilibrium, the condition is 
\begin{equation}\label{cond2}
  \frac{d^2 S}{dx^2} < 0. 
 \end{equation}

\begin{figure*} 
	\begin{center}
		\includegraphics[angle=0, width=\textwidth]{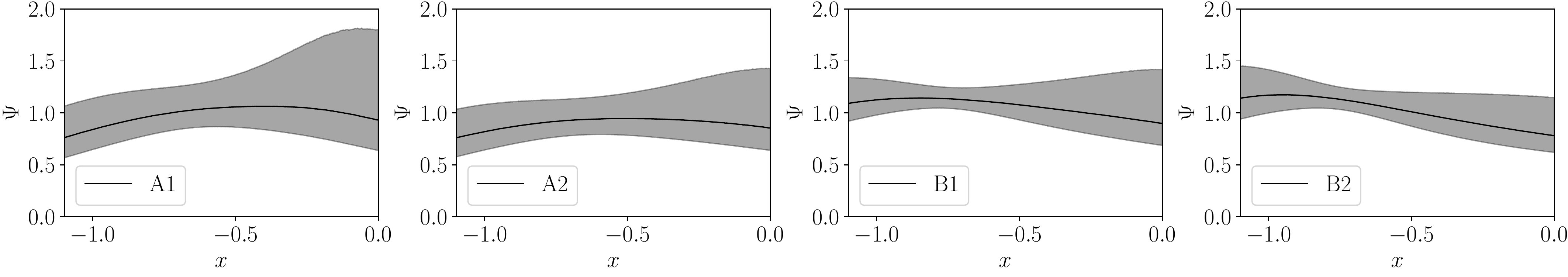}\\
		\includegraphics[angle=0, width=\textwidth]{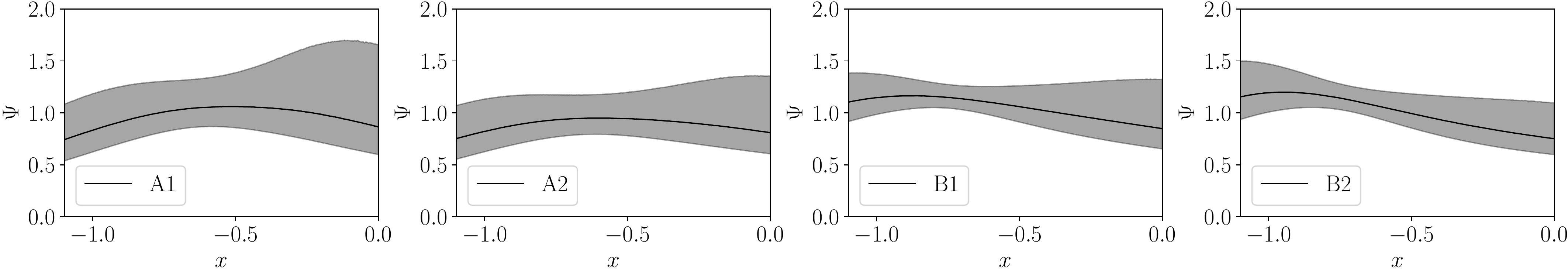}
	\end{center}
	\caption{{\small Plots for $\Psi$ from the dataset samples of Set A and B using a Squared Exponential covariance (top row) and the Mat\'{e}rn $9/2$ covariance 
			function. The solid black line gives the best fit values of $\Psi$. The shaded region correspond to the 1$\sigma$ uncertainty.}}
	\label{psi_plot}
\end{figure*}

\begin{figure*} [t!]
	\begin{center}
		\includegraphics[angle=0, width=\textwidth]{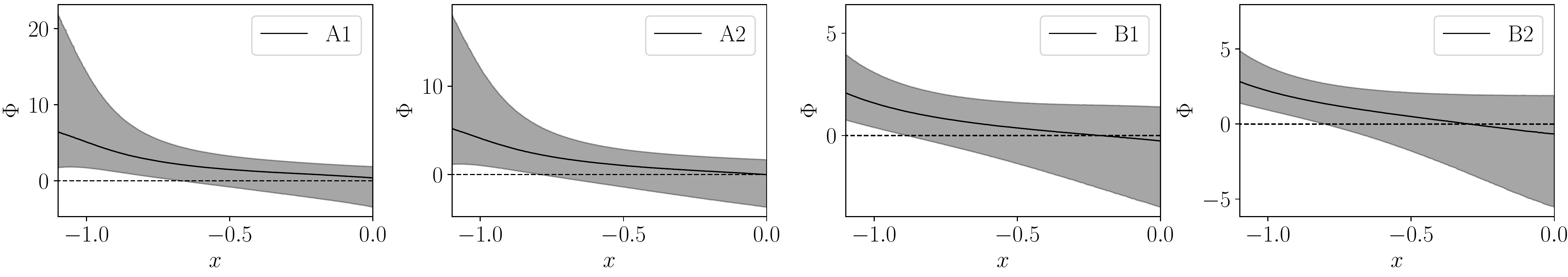}\\
		\includegraphics[angle=0, width=\textwidth]{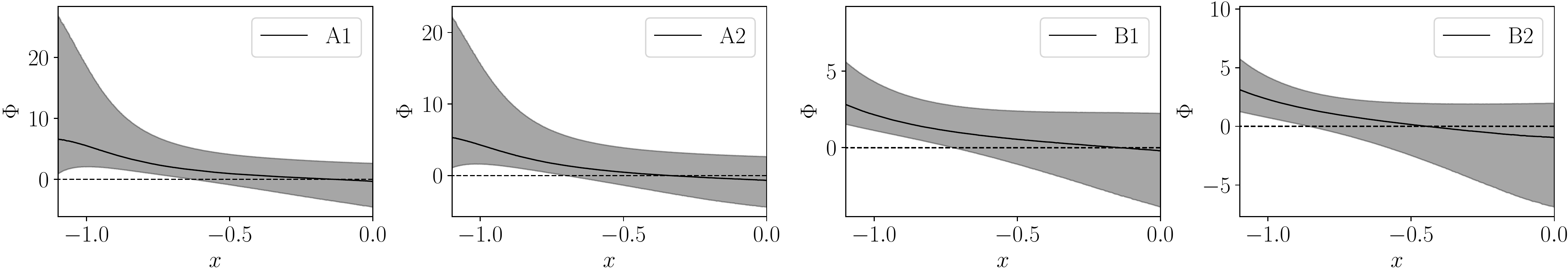}
	\end{center}
	\caption{{\small Plots for $\Phi$ from the dataset samples of Set A and B using a Squared Exponential covariance (top row) and the Mat\'{e}rn $9/2$ covariance 
			function. The solid black line gives the best fit values of $\Phi$. The shaded region correspond to the 1$\sigma$ uncertainty.}}
	\label{phi_plot}
\end{figure*}

With the apparent horizon as the cosmological horizon, the entropy of the horizon $S_h$ can be written as \cite{bak}
\begin{equation} \label{Sh}
	S_h =  8 \pi^2 r_h^2 = \frac{8 \pi^2}{H^2}.
\end{equation}
Further, the temperature of the dynamical apparent horizon is related to the horizon radius by the relation,
\begin{equation} \label{HKtemp}
	T_h = \frac{1}{2 \pi r_h} \left[ 1 - \frac{\dot{r_h}}{2 H r_h}\right] = \frac{2 H^2 + \dot{H}}{4 \pi H}~,
\end{equation} 
which is called the Hayward-Kodama temperature \cite{hktemp1, hktemp2}. As the cosmological horizon is evolving, Hawking temperature is replaced 
by Hayward-Kodama temperature (see \cite{valerio}). \\

Now, if we denote $S_f = S_m + S_D$, where $S_m$, $S_D$ being the 
entropies of the matter sector and the dark energy sector, with $T$ as the temperature of composite matter distribution inside the horizon, then the first law of 
thermodynamics $TdS = dE + p~dV$ can be recast for the individual matter components in the following form, 
\begin{eqnarray} 
T~dS_m &=& dE_m + p_m dV = dE_m , \label{tdsm} \\
T~dS_D &=& dE_D + p_D dV, \label{tdsd}
\end{eqnarray} 
where $V = \frac{4}{3} \pi r_h^3 = \frac{4 \pi}{3 H^2}$, is the fluid volume. $E_m$, $E_D$ represent the internal energies of the dark matter 
and energy components given by $E_m = \frac{4}{3}\pi r_h^3 \rho_{m} = \rho_{m} V$ and $E_D = \frac{4}{3}\pi r_h^3 \rho_{D} = \rho_{D} V$ respectively. Now, 
differentiating equations \eqref{Sh}, \eqref{tdsm} and \eqref{tdsd} w.r.t. the cosmic time $t$ along with the assumption $T$ should be equal to 
$T_h$ (equation \eqref{HKtemp}) we get, \\
\begin{eqnarray}
	\dot{S_m} + \dot{S_D} &=& 16 \pi^2 \frac{\dot{H}}{H^3} \left(1+\frac{\dot{H}}{2H^2+\dot{H}}\right), \\
	\dot{S_h} &=& -16 \pi^2 \frac{\dot{H}}{H^3}.
\end{eqnarray}

Therefore,
\begin{equation} \label{Sdot}
	\dot{S} = \dot{S_m} + \dot{S_D} +\dot{S_h} = 16 \pi^2 \frac{\dot{H}^2}{H^3} \left(\frac{1}{2H^2+\dot{H}}\right).
\end{equation}
One should note that it may not always be justified to assume the fluid temperature to be equal to the horizon temperature. This assumption is 
particularly unjustified for a radiation distribution that obeys Stefan's law. However, for a pressureless matter, the equality of $T$ and $T_h$ 
is valid and for dark energy, this equality is at least approximately correct. Thus in the present context, this assumption is not at all 
drastic. For an account of this justification, we refer to the work of Mimoso and Pav\'{o}n\cite{mimoso}. \\

The relation \eqref{Sdot} can be written with $x$ as the argument, where $x = \ln ~a = -\ln (1+z)$, as 
\begin{equation} \label{Sdiffx}
\frac{d S}{d x} = \frac{16 \pi^2}{H^4} \left(\frac{dH}{dx}\right)^2 \Psi(x),
\end{equation} 
where 
\begin{equation}
	\Psi(x) = \left[2 + \frac{1}{H} \frac{dH}{dx}\right]^{-1}.
\end{equation}

Again on differentiating equation \eqref{Sdiffx} w.r.t. $x$, one obtains,
\begin{equation} \label{Sdiff2x}
	\frac{d^2 S}{d x^2} = \frac{16 \pi^2 \Psi^2}{H^4} \left( \frac{dH}{dx}\right)^2 \Phi(x),
\end{equation} 
where, 
\begin{equation}
	\Phi = \frac{1}{H}\frac{d^2H}{dx^2} - \frac{3}{H^2}\left(\frac{dH}{dx}\right)^2 + \frac{4}{\frac{dH}{dx}} \frac{d^2H}{dx^2} - \frac{8}{H} \frac{dH}{dx} .
\end{equation}

From equation \eqref{Sdiffx}  we see that for the inequality (\ref{cond1}) to hold true, the required condition is $\Psi \geq 0$. Equation \eqref{Sdiff2x} shows that 
condition (\ref{cond2}) shall 
be satisfied, provided $\Phi<0$. We try to understand the behaviour of $\Psi$ and $\Phi$ by plotting them, in figure \ref{psi_plot} and \ref{phi_plot} respectively,  as 
functions of $x$ where $x$ = $-\ln(1+z)$. \\

The plots in figure \ref{psi_plot} show that $\Psi$ remains positive in 1$\sigma$ throughout the domain of reconstruction $0<z<2$. Thus, the second law of thermodynamics 
is indeed satisfied for the reconstructed scenario. From figure \ref{phi_plot} the plots reveal that $\Phi$ was  positive in the past but as we approach the present epoch 
the value of $\Phi$ decreases gradually and changes its signature. The best fit value $\Phi$ becomes negative as $x$ increases. This hints towards a possibility that 
the Universe is undergoing a change from a thermodynamic non-equilibrium in the past towards an equilibrium state in the present epoch. \\

\section{Discussion}

It is argued quite often that the possibility of a non-gravitational interaction in the cosmic dark sector should not be ruled out \textit{a priori}. As the nature of dark 
energy is not known, it is impossible to model the interaction theoretically. The normal practice is to assume a transfer of energy between the dark matter 
and the dark energy and write the rate of transfer $Q$ as a function of the densities $\rho_D$, or $\rho_m$, or both, and even their derivatives\cite{yangnbpan} and thus 
parametrize $Q$. The next step is to reconstruct the model parameters using the observational data. This is indeed biased in a way, as the functional form of $Q$ is 
already chosen. \\

The present work employs the widely practised Gaussian Process and makes an attempt to reconstruct the transfer of energy $Q$ in a dimensionless representation (defined as 
$\tilde{Q} = \frac{Q}{3H_{0}^{2}}$) directly from observational datasets without any parametrization. Thus no functional form of $Q$ is assumed. Various combinations of 
datasets are utilized, properly described in section III. There have been only a few investigations on a non-parametric reconstruction of $Q$ as mentioned in the 
introduction. The primary difference of the present work with the existing literature is that we used more recent datasets and investigated for three different versions of 
dark energy. They are (i) an interacting vacuum with $w=-1$, (ii) a $w$CDM model where $w$ is close to $-1$ but not exactly equal to that and (iii) the CPL parametrization 
where $w(z) = w_0 + w_a \frac{z}{1+z}$. The universal feature that we find is that for any of these choices and any combination of datasets, an interaction in the dark sector 
is not significant at the present epoch. The interaction may not be ruled out in the past, beyond $z\geq 0.5$, but a zero $Q$ is indeed a possibility 
normally in $2\sigma$ and at most in $3\sigma$. This result is different from the  oscillatory behaviour as noted by Cai and Su\cite{cai_su}. Our result is closer to 
that given by  Wang \textit{et al.}\cite{wang_pca} where a non-parametric Bayesian approach indicated that an interacting vacuum is not preferred. \\

An analytic expression for the energy transfer rate $Q$ in the form of a polynomial in $z$ is given in section IV. The reduced ${\chi}^2$ test allowed upto second 
order in $z$ for some combinations (A1 and A2) while only upto first order in $z$ for the other two combinations (B1 and B2) of datasets. \\

Evolution of the density parameters ${\Omega}_m$ and ${\Omega}_D$ are also checked in the presence of the interaction in the dark sector. The common feature that comes out is that 
the dominance of dark energy is delayed a bit (closer to $z=0$).\\

The thermodynamic considerations reveal an interesting possibility. While the reconstructed interaction does not infringe upon the thermodynamic viability in terms of the 
increase in entropy, the universe seems to be evolving towards a thermodynamic equilibrium only from a recent past, namely $x\sim -0.5$, i.e., close to $z\sim 0.6$.\\


\textbf{Acknowledgement}\\

PM thanks her colleagues for lively discussions.

\end{document}